%% file: eye-computations-amsart.tex
\documentclass[reqno,a4paper]{amsart}
\usepackage{amsmath}
\usepackage{amsthm}
\usepackage{amssymb}

\usepackage[scale=0.9]{geometry}

\usepackage[numbers]{natbib}

\usepackage[english]{babel}
\usepackage[utf8]{inputenc}

\usepackage{subfig}
\usepackage{graphicx}

\usepackage[unicode,breaklinks]{hyperref}

\usepackage{mathbbol}
\usepackage{bm} 
\usepackage{gensymb}

\usepackage{units}
\usepackage{tensor}
\usepackage{accents}

\usepackage{booktabs}
\usepackage{threeparttable}

\usepackage[section]{placeins} 

\newlength{\scolstep} 
\newlength{\scoldesc} 
\input{vit-prusa-macros-experimental}

\newcommand{\completedeformation}{\greekvec{\varphi}}
\newcommand{\displacementu}{\vec{u}}
\newcommand{\displacementucomp}{u}

\let\cite\citet
\numberwithin{equation}{section}

\makeatletter
\providecommand{\doi}[1]{%
  \begingroup
    \let\bibinfo\@secondoftwo
    \urlstyle{rm}%
    \href{http://dx.doi.org/#1}{%
      doi:\discretionary{}{}{}%
      \nolinkurl{#1}%
    }%
  \endgroup
}
\makeatother

\author{Karel T\r{u}ma}
\address{%
Faculty of Mathematics and Physics, Charles University\\
Sokolovsk\'a 83, Praha 8 -- Karl\'{\i}n, CZ 186\;75, Czech Republic}
\email{ktuma@karlin.mff.cuni.cz}

\author[hdl]{Judith Stein} 
\address{%
Institute of Applied Mathematics, Faculty of Mathematics and Computer Sciences, Heidelberg University\\
Im~Neuenheimer Feld 205, Heidelberg, DE 69120, Germany}
\email{judith.stein@iwr.uni-heidelberg.de}

\author[cuni]{V\'{\i}t Pr\r{u}\v{s}a}
\address{%
Faculty of Mathematics and Physics, Charles University\\
Sokolovsk\'a 83, Praha 8 -- Karl\'{\i}n, CZ 186\;75, Czech Republic}
\email{prusv@karlin.mff.cuni.cz}

\author[hdl]{Elfriede Friedmann}
\address{%
Institute of Applied Mathematics, Faculty of Mathematics and Computer Sciences, Heidelberg University\\
Im~Neuenheimer Feld 205, Heidelberg, DE 69120, Germany}
\email{friedmann@iwr.uni-heidelberg.de}

\thanks{Karel T\r{u}ma and V\'{\i}t Pr\r{u}\v{s}a acknowledge the support of the Project LL1202 in the programme ERC-CZ funded by the Ministry of Education, Youth and Sports of the Czech Republic. Karel T\r{u}ma has been supported by Charles University Research program No.~UNCE/SCI/023. The financial support of Judith Stein and Elfriede Freidmann for the project No.00.265.2015 is provided by the Klaus Tschira Stiftung gGmbH.}

\title{Motion of the vitreous humour in a deforming eye -- fluid-structure interaction between a nonlinear elastic solid and a nonlinear viscoleastic fluid}

\keywords{viscoelastic fluid, Burgers model, nonlinear elasticity, fluid structure interaction, eye, vitreous, sclera}

\subjclass[2000]{ 76A05, 
74F10, 
92C10, 
76Z05 
}

\begin{document}

\begin{abstract}
  \input{eye-computations-abstract}
\end{abstract}

\maketitle

\section{Introduction}
\label{sec:introduction}

\input{introduction}

\section{Outline}
\label{sec:outline}

\input{outline}

\section{Problem description}
\label{sec:problem-description}

\input{problem}

\section{Constitutive relations}
\label{sec:const-relat}

\input{constitutive}

\section{Full system of governing equations}
\label{sec:full-syst-govern}

\input{governing}

\section{Numerical solution of the fluid-structure interaction problem}
\label{sec:numer-solut-fluid}

\input{numerics}

\section{Results}
\label{sec:results}

\input{results}

\section{Conclusion}
\label{sec:conclusion}

\input{conclusion}

\bigskip

\bibliographystyle{elsarticle-num-names}
\bibliography{vit-prusa,literature}
\end{document}

%% file: vit-prusa-macros-experimental.tex
\DeclareMathOperator{\divergence}{div}
\DeclareMathOperator{\Divergence}{Div}

\DeclareMathOperator{\Tr}{Tr}

\newcommand{\bydefinition}{\mathrm{def}}

\newcommand{\diff}{\mathrm{d}}

\renewcommand{\vec}[1]{\ensuremath{\mathbf{#1}}}
\newcommand{\greekvec}[1]{\ensuremath{\boldsymbol{#1}}}
\makeatletter
\@ifpackageloaded{bm}%
{\renewcommand{\vec}[1]{\ensuremath{\bm{#1}}}%
\renewcommand{\greekvec}[1]{\ensuremath{\bm{#1}}}%
}{%
\relax
}
\makeatother
\newcommand{\tensorq}[1]{\ensuremath{\mathbb{#1}}}      

\newcommand{\transpose}[1]{#1^\top}

\newcommand{\identity}{\ensuremath{\tensorq{I}}}

\newcommand{\cstress}{\tensorq{T}}

\newcommand{\ecstress}{\tensorq{S}}

\newcommand{\fgrad}{\tensorq{F}}

\newcommand{\rcg}{\tensorq{C}}

\newcommand{\lcg}{\tensorq{B}}

\makeatletter
\@ifpackageloaded{bm}%
{%
}{%

}

\@ifpackageloaded{bm}%
{%
 
}{%

}

\@ifpackageloaded{bm}%
{%
}{%

}
\makeatother

\newcommand{\generictensor}{{\tensorq{A}}}

\newcommand{\vecv}{\ensuremath{\vec{v}}}
\newcommand{\gradv}{\ensuremath{\nabla \vecv}}

\newcommand{\gradsym}{\ensuremath{\tensorq{D}}}

\newcommand{\bvec}[1]{\vec{e}_{#1}} 

\newcommand{\bvecx}{\bvec{\hat{x}}}
\newcommand{\bvecy}{\bvec{\hat{y}}}
\newcommand{\bvecz}{\bvec{\hat{z}}}

\newcommand{\pd}[2]{\ensuremath{\frac{\partial {#1}}{\partial {#2}}}}

\newcommand{\fid}[1]{\ensuremath{\accentset{\triangledown}{#1}}}
\newcommand{\fidd}[1]{\ensuremath{\accentset{\triangledown \! \triangledown}{#1}}}

\makeatletter
\@ifpackageloaded{tensor}
{

}{%

}
\makeatother

\makeatletter
\@ifpackageloaded{tensor}
{

}{%

}
\makeatother

\makeatletter
\@ifundefined{volume}{%
}%
{%
}
\makeatother

\newcommand{\surfacees}{\diff \mathrm{S}}

\makeatletter
\@ifpackageloaded{MnSymbol} 
{
 
}{%
 
}
\makeatother
\newcommand{\vectordot}[2]{\ensuremath{#1 \bullet #2}}



%% file: eye-computations-abstract.tex
We study the motion of vitreous humour in a deforming eyeball. From the mechanical and computational perspective this is a task to solve a fluid-structure interaction problem between a complex viscoelastic fluid (vitreour humour) and a nonlinear elastic solid (sclera and lens). We propose a numerical methodology capable of handling the fluid-structure interaction problem, and we demonstrate its applicability via solving the corresponding governing equations in a realistic geometrical setting and for realistic parameter values. It is shown that the choice of the rheological model for the vitreous humour has a negligible influence on the overall flow pattern in the domain of interest, whilst it is has a significant impact on the mechanical stress distribution in the domain of interest.


%% file: introduction.tex
The vitreous humour is a fluid like material that occupies the space between the lens and the retina in the eyeball. It has several functions, some of them being of mechanical nature. In particular, the vitreous humour is essential in the transmission of the mechanical stresses in the eyeball, and it acts as a mechanical damper protecting the eye, see for example~\cite{kleinberg.tt.tzekov.rt.ea:vitreous} and references therein. Consequently, the \emph{mechanical properties} of the vitreous humour and its \emph{motion} are important both in the understanding of the physiology and pathology of the eye. 

The motion of the vitreous humour can be induced either by motion of the eyeball as the whole or by the deformation of the eyeball. In the latter case the vitreous humour interacts with the deforming sclera and the lens, which from the mechanical point of view constitutes a complicated fluid-structure interaction problem. The complexity of the problem rests in the fact that one needs to deal with an interaction between a non-Newtonian viscoelastic fluid (vitreous humour) and a nonlinear elastic solid (sclera and lens). 

We present a numerical methodology that is capable of handling the fluid-structure interaction problem.  Then we apply the methodology in numerically solving the governing equations in a setting that resembles the recent experiment by~\cite{shah.ns.beebe.dc.ea:on}. The numerical experiment shows that the adopted methodology can be used to predict key mechanical quantities such as the mechanical stress at the interface between the vitreous humour and the retina. These quantities are of interest in study of several pathologies such as retinal detachment, which documents the usability of the proposed methodology in answering practically relevant questions. 

The main contribution of the current study is twofold. First, the majority of the works focused on the motion of the vitreous humor have been so far focused on the saccadic eye movement (rapid oscillations of the eyeball as the whole), see the early study by~\cite{david.t.smye.s.ea:model} and numerous later studies by \cite{repetto.r:analytical}, \cite{repetto.r.siggers.jh.ea:mathematical}, \cite{meskauskas.j.repetto.r.ea:oscillatory}, \cite{meskauskas.j.repetto.r.ea:shape}, \cite{bonfiglio.a.repetto.r.ea:investigation}, \cite{abouali.o.modareszadeh.a.ea:numerical}, \cite{modarreszadeh.a.abouali.o:numerical} and \cite{repetto.r.siggers.jh.ea:steady} to name a few. In all these works, the geometry of the cavity filled by the vitreous humour is assumed to be \emph{fixed}. In particular, the cavity is assumed to take either the spherical shape or a perturbed spherical shape. This is perfectly acceptable if the motion of the vitreous humour is \emph{induced by the oscillation of the eyeball as the whole}, and important conclusion has been drawn in the referred works. However, if the motion of the vitreous humour is \emph{induced by the deformation of the sclera}, then a completely different approach must be taken. The problem must be solved as a fluid-structure interaction between the deforming sclera/lens and the flowing vitreous humour.

Second, the early works focused on the motion of the vitreous humor predominantly assume that the vitreous humour is an incompressible Navier--Stokes fluid. This is a plausible assumption especially if the motion of pathological (liquefied) vitreous humour is of interest. The physiological vitreous humour is however known to exhibit viscoelastic properties, see the early study by~\cite{lee.b.litt.m.ea:rheology} or a more recent study by~\cite{sharif-kashani.p.hubschman.j.ea:rheology}. Consequently, the impact of complex viscoelastic (non-Newtonian) rheology on the motion of the vitreous humour must be taken into account. Various relatively simple viscoelastic models have been recently studied in this respect. For example, \cite{modarreszadeh.a.abouali.o:numerical} in their numerical study use the nonlinear viscoelastic model introduced (for polymeric fluids) by~\cite{giesekus.h:simple}. On the other hand, yet more complex models have been introduced in order to fit the experimental data. In particular, \cite{sharif-kashani.p.hubschman.j.ea:rheology} have described the mechanical response of the vitreous humour using a Burgers-type viscoelastic model, see~\cite{burgers.jm:mechanical}. So far, the vitreous humour motion predicted by this advanced model has been investigated neither by analytical nor by numerical methods.   

In what follows we address both issues. First, we consider flow of the vitreous humour in a \emph{deforming cavity}. The deformation of the cavity is induced by the deformation of the sclera, while the deformation of the sclera is caused by an applied load. Second, the mechanical properties of the vitreous humour are in the present study described by a relatively complex \emph{Burgers-type viscoleastic rate-type model}. Moreover, since we need to solve the problem for the deformation of the sclera, we also need a model for the response of this solid substance. Concerning the mechanical response of the sclera, we assume that it behaves as a (nonlinear) \emph{hyperelastic solid}. 


%% file: outline.tex
The mathematical models used for the description of the response of the vitreous humour and sclera are introduced in detail in Section~\ref{sec:const-relat}. Note that since the experimental data for the vitreous humour are interpreted using \emph{one-dimensional} spring-dashpot analogues, see~\cite{sharif-kashani.p.hubschman.j.ea:rheology}, we need to provide a three-dimensional variant of the one-dimensional model. This step is not a straightforward one, and in addressing this issue we follow the thermodynamics based approach introduced by~\cite{rajagopal.kr.srinivasa.ar:thermodynamic}.

The numerical methodology for the solution of the arising fluid-structure interaction problem is discussed in Section~\ref{sec:numer-solut-fluid}. The numerical methodology is in principle based on the arbitrary Lagrangian--Eulerian method, see for example~\cite{donea.j.huerta.a.ea:arbitrary}.

Finally, the proposed numerical method is used to solve the governing equations in the geometrical setting that resembles the recent experimental setting studied by~\cite{shah.ns.beebe.dc.ea:on}. In particular, see Section~\ref{sec:results}, we focus on comparison of the flow fields and the mechanical stress distributions obtained in two scenarios. In the first scenario we assume that the mechanical response of the vitreous humour is described by the standard Navier--Stokes model, while in the second scenario the vitreous humour is described by the Burgers-type viscoelastic model. It is shown that choice of the rheological model for the vitreous humour has significant impact on the mechanical stress distribution in the domain of interest, whilst the overall flow field remains almost independent on the choice of the rheological model.  


%% file: problem.tex

\begin{figure}[h]
 \centering	 
    \includegraphics[width=12cm]{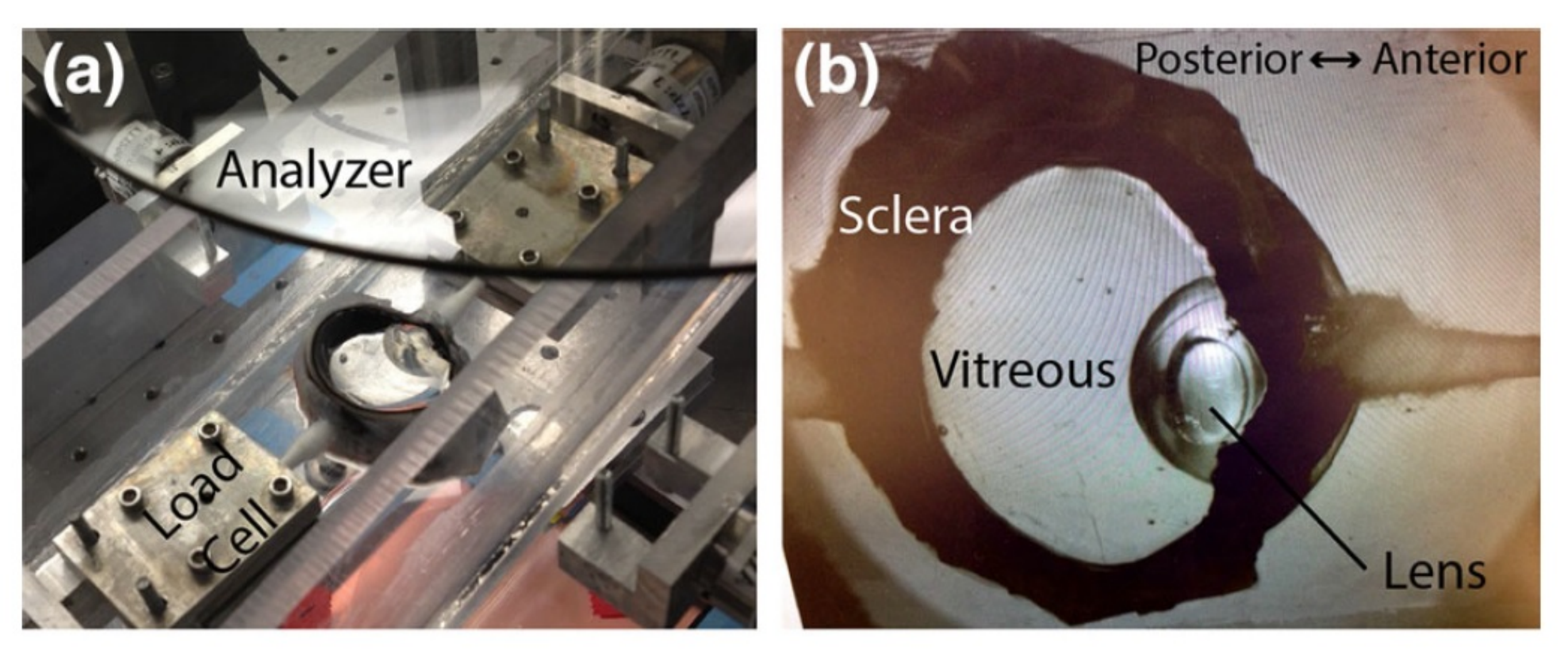}
 \caption{Experiment by~\cite{shah.ns.beebe.dc.ea:on}. (Reprinted by permission from Springer Customer Service Centre GmbH: Springer Nature, Annals of Biomedical Engineering, \citeauthor{shah.ns.beebe.dc.ea:on}, On the spatiotemporal material anisotropy of the vitreous body in tension and compression, \textcopyright\ 2016.)}
  \label{fig:experiment}
\end{figure}

\begin{figure}[h]
\begin{center}
\includegraphics[width=6cm]{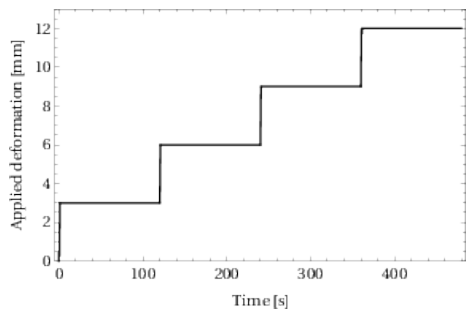}
\end{center}
\caption{Prescribed deformation.}
\label{fig: Prescribed deformation}
\end{figure}

In the experiments from \cite{shah.ns.beebe.dc.ea:on} fresh bovine eyes were cut in an anterior-posterior direction to create approximately \unit[2]{cm} thick samples with an optically clear window to analyze the changes during the experiment, see Figure \ref{fig:experiment}. Placed in an anterior-posterior orientation to the load cells the samples were attached at the sclera and near the lens by cotton swabs fixed via clamps to the load cells, Figure \ref{fig:experiment}. Then the samples were uniaxially stretched by simultaneously moving both load cells in \unit[3]{mm} increments (up to \unit[12]{mm}) with \unit[120]{s} of equilibration time between each loading step, see Figure~\ref{fig: Prescribed deformation}. Mechanical strain was measured from sparse marker arrays on the surface of the vitreous and temporal collagen behavior was estimated from creep compliance rheological tests.

\subsection{Geometry}
\label{sec:geometry}
\begin{figure}[h]
 \centering	 
    \includegraphics[width=0.2\textwidth]{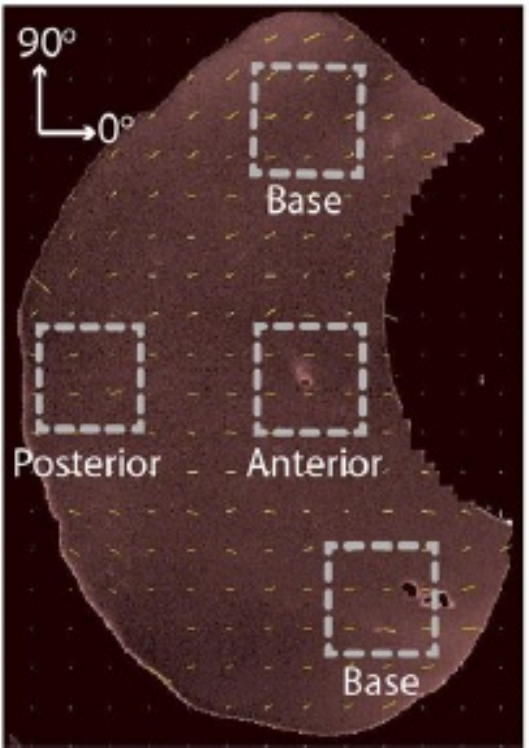}
 \caption{Undeformed vitreous, \cite{shah.ns.beebe.dc.ea:on}. (Reprinted by permission from Springer Customer Service Centre GmbH: Springer Nature, Annals of Biomedical Engineering, \citeauthor{shah.ns.beebe.dc.ea:on}, On the spatiotemporal material anisotropy of the vitreous body in tension and compression, \textcopyright\ 2016.)}
  \label{fig:undeformed-ex}
\end{figure}

\begin{figure}[h]
  \centering
  \subfloat[\label{fig:geometry-a}Domain decomposition in the reference configuration -- lens, sclera and vitreous humour.]{\includegraphics[height=0.35\textwidth]{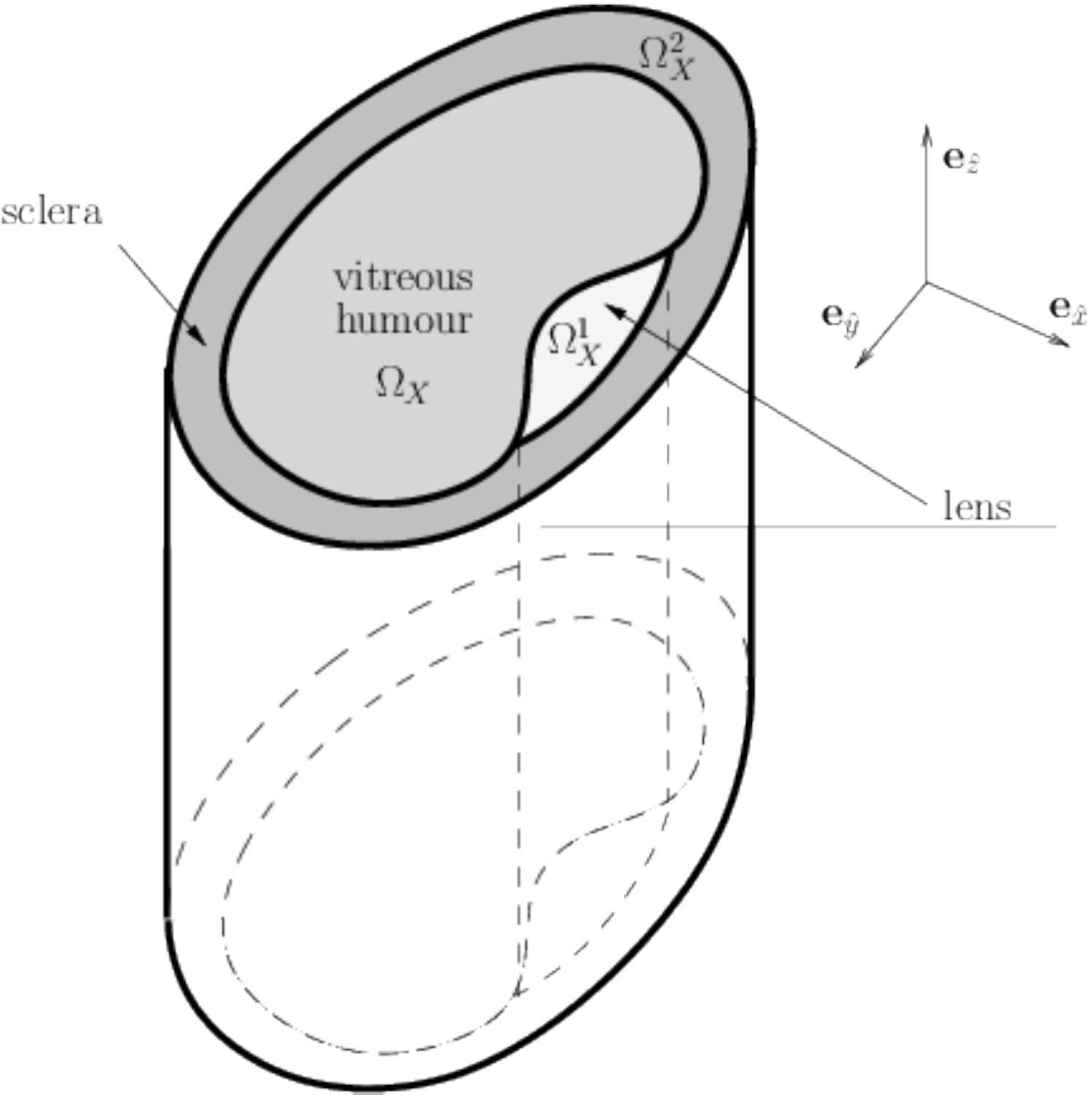}}
  \qquad
  \subfloat[\label{fig:geometry-b}Surface decomposition in the reference configuration -- boundary conditions.]{\includegraphics[height=0.35\textwidth]{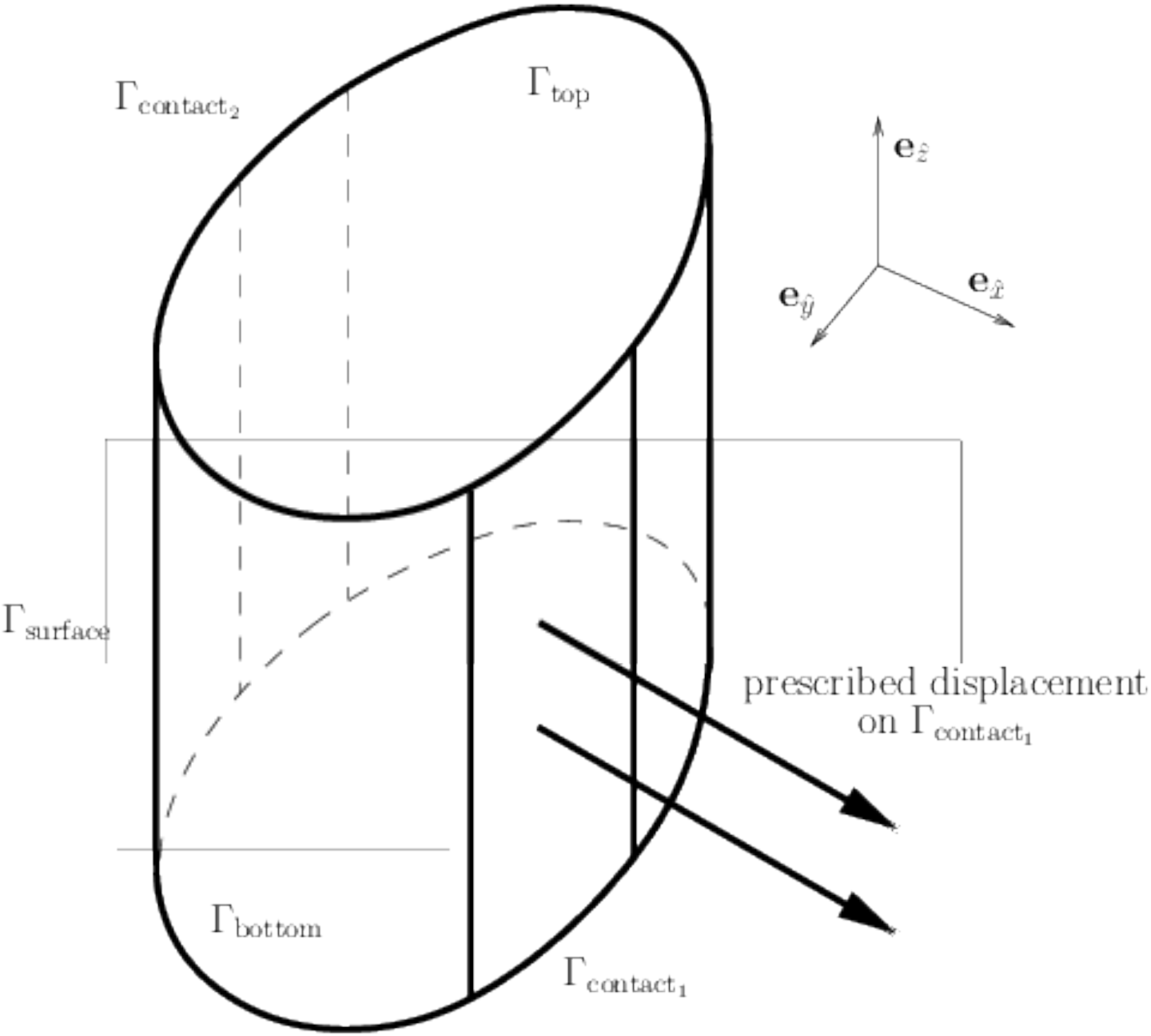}}
  \\
  \subfloat[\label{fig:mesh}Discretization of the reference configuration using the \textsc{Gmsh} mesh generator, see~\cite{geuzaine.c.remacle.j:gmsh};  vitreous humour (red), sclera (brown), lens (yellow). Complete quantitative description of the geometry is available in the supplementary material. 
  ]{\includegraphics[width=0.3\textwidth]{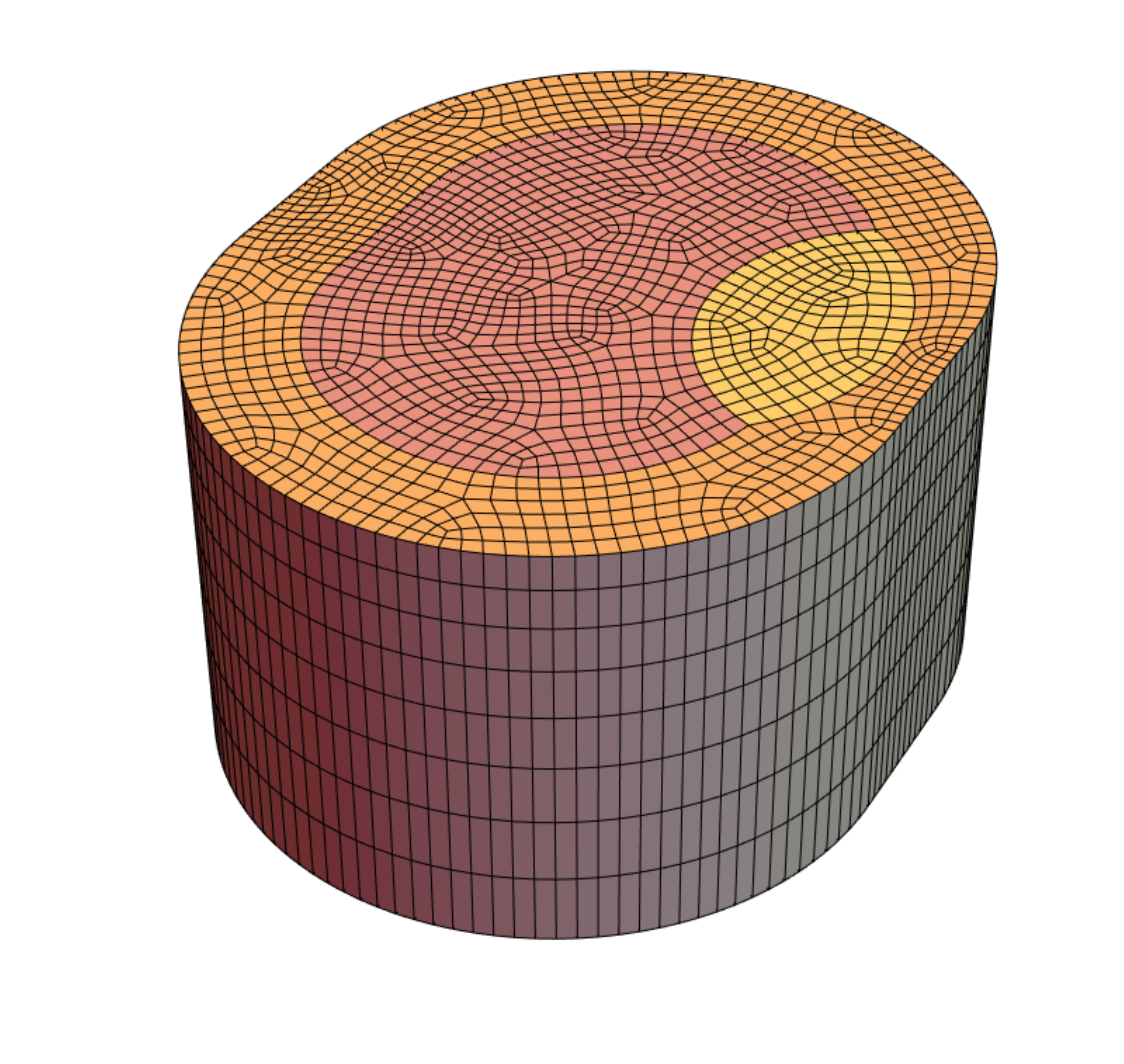}}
  \caption{Problem geometry.}
  \label{fig:geometry}
\end{figure}

Motivated by these experiments we consider the following problem. The sample is modelled as a \unit[2]{cm} high cylinder, see Figure~\ref{fig:geometry-a}, while the cylinder cross-section has the shape similar to that in the experiment by~\cite{shah.ns.beebe.dc.ea:on}, see Figure~\ref{fig:undeformed-ex}. The domain is composed of three parts, the vitreous humor occupies the cavity $\Omega_X$ that is formed by the lens $\Omega_X^1$ and the sclera~$\Omega_X^2$, see Figure~\ref{fig:geometry-a}. The thickness of the sclera is around \unit[5]{mm} around the vitreous and the lens, the exact dimensions used for the mesh generation for the finite element method are available as a supplementary material. 

\subsection{Boundary conditions}
\label{sec:boundary-conditions}

Concerning the boundary conditions, we prescribe boundary conditions that again resemble the experimental setting by~\cite{shah.ns.beebe.dc.ea:on}. The bottom base of the cylinder $\Gamma_{\mathrm{bottom}}$ that is placed at $z=0$, see Figure~\ref{fig:geometry-b}, is assumed to be immovable in the direction of the $z$-axis, hence the $z$ component of the displacement vector $\vec{u}$ is fixed to zero. Further, the bottom base is allowed to move freely in the horizontal direction, hence in the direction of $x$ and $y$-axis we prescribe no-traction boundary conditions on $\completedeformation(\Gamma_{\mathrm{bottom}})$, where $\completedeformation$ denotes the deformation between the reference and current configuration. In total, on $\Gamma_{\mathrm{bottom}}$ we set
\begin{subequations}
  \label{eq:boundary-conditions}
  \begin{align}
    \label{eq:2}
    \left. \vectordot{\displacementu}{\bvecz} \right|_{\completedeformation(\Gamma_{\mathrm{bottom}})} &= 0, \\
    \label{eq:3}
    \left. \vectordot{\cstress \vec{n}}{\bvecx} \right|_{\completedeformation(\Gamma_{\mathrm{bottom}})} &= 0, \\
    \label{eq:4}
    \left. \vectordot{\cstress \vec{n}}{\bvecy} \right|_{\completedeformation(\Gamma_{\mathrm{bottom}})} &= 0, 
  \end{align}
  where $\cstress$ denotes the Cauchy stress tensor, and $\vec{n}$ denotes the unit outward normal in the current configuration. The top base of the cylinder $\Gamma_{\mathrm{top}}$ is assumed to be traction free, and we set
  \begin{equation}
    \label{eq:5}
    \left. \cstress \vec{n} \right|_{\completedeformation(\Gamma_{\mathrm{top}})} = \vec{0},
  \end{equation}
  and the same holds for the lateral walls of the cylinder,
  \begin{equation}
    \label{eq:6}
    \left. \cstress \vec{n} \right|_{\completedeformation(\Gamma_{\mathrm{surface}})} = \vec{0}.
  \end{equation}
  On the vertical strips $\Gamma_{\mathrm{contact}_1}$ and $\Gamma_{\mathrm{contact}_2}$ we prescribe the displacement corresponding to the experiment by~\cite{shah.ns.beebe.dc.ea:on}. Both on the anterior part $\Gamma_{\mathrm{contact}_1}$ and the posterior part $\Gamma_{\mathrm{contact}_2}$ we prescribe the displacement in the $x$-axis direction $\displacementucomp_x =_{\bydefinition} \vectordot{\displacementu}{\bvecx}$ and we allow the sample to freely move in the other directions
  \begin{align}
    \label{eq:7}
    \left. \vectordot{\cstress \vec{n}}{\bvecy} \right|_{\completedeformation(\Gamma_{\mathrm{contact}_{12}})} &= 0, \\
    \label{eq:4}
    \left. \vectordot{\cstress \vec{n}}{\bvecz} \right|_{\completedeformation(\Gamma_{\mathrm{contact}_{12}})} &= 0.
  \end{align}
On the posterior part $\Gamma_{\mathrm{contact}_2}$ we fix the displacement in the $x$-axis direction $\displacementucomp_x=0$. The displacement $\displacementucomp_x$ on the anterior part $\Gamma_{\mathrm{contact}_1}$ is increased in four prolongation steps by $\unit[3]{mm}$. One prolongation step takes one second with the corresponding velocity of \unitfrac[3]{mm}{s}. Together with \unit[119]{s} to relax afterwards each prolongation step we have \unit[2]{min} of equilibration time between each displacement step, see Figure~\ref{fig: Prescribed deformation} for the plot of the prescribed displacement versus time.
\end{subequations}
The exact quantitative description of locations of the surface components $\Gamma_{\mathrm{bottom}}$, $\Gamma_{\mathrm{top}}$, $\Gamma_{\mathrm{surface}}$ and $\Gamma_{\mathrm{contact}_1}$ and $\Gamma_{\mathrm{contact}_2}$ is given in the supplementary material. 


%% file: constitutive.tex

\subsection{Vitreous humour}
\label{sec:vitreous-humour}

The available experimental data, see \cite{sharif-kashani.p.hubschman.j.ea:rheology}, indicate that the rheological behaviour of the vitreous humour can be described by an incompressible viscoelastic Burgers-type model, whilst the usage of a complex rheological model is enforced by the presence of multiple relaxation mechanisms. The Burgers-type model is, as many viscoelastic rate-type models, see for example~\cite{wineman.as.rajagopal.kr:mechanical}, motivated by a one-dimensional mechanical spring/dashpot analogue. The one-dimensional mechanical analogue consists of two linear dashpots and two linear springs with an additional dashpot to obtain a more convenient (from the perspective of fitting the experimental data) model than the original Burgers model.

\begin{figure}[h]
 \centering	 
 \subfloat[\label{fig:karel}Two Maxwell elements with an additional dashpot in parallel.]{\includegraphics[height=0.25\textwidth]{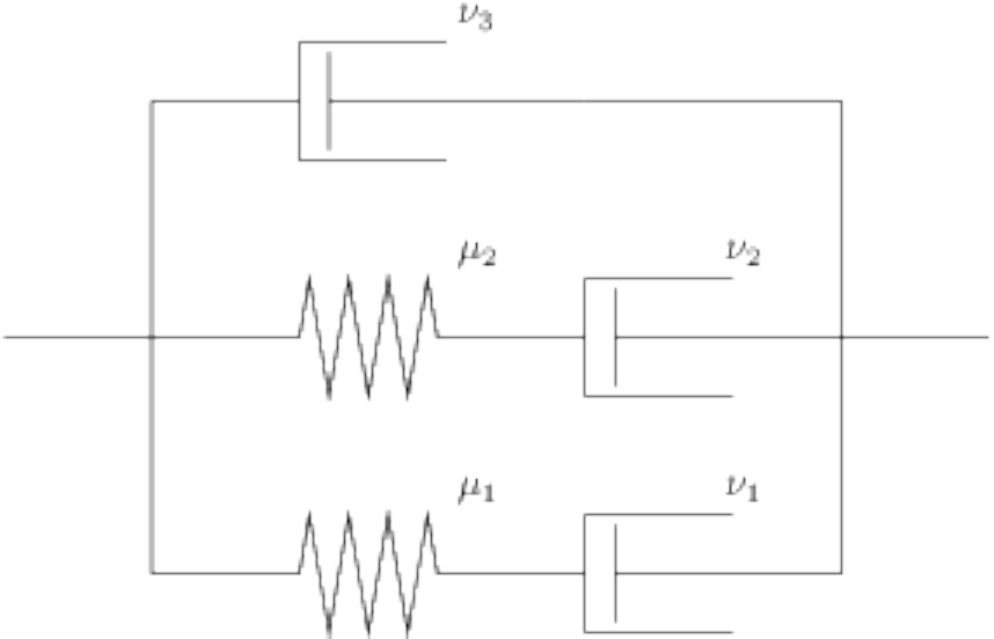}}
 \qquad
 \subfloat[\label{fig:paper}Two Kelvin--Voigt elements with an additional dashpot in series.]{\includegraphics[height=0.25\textwidth]{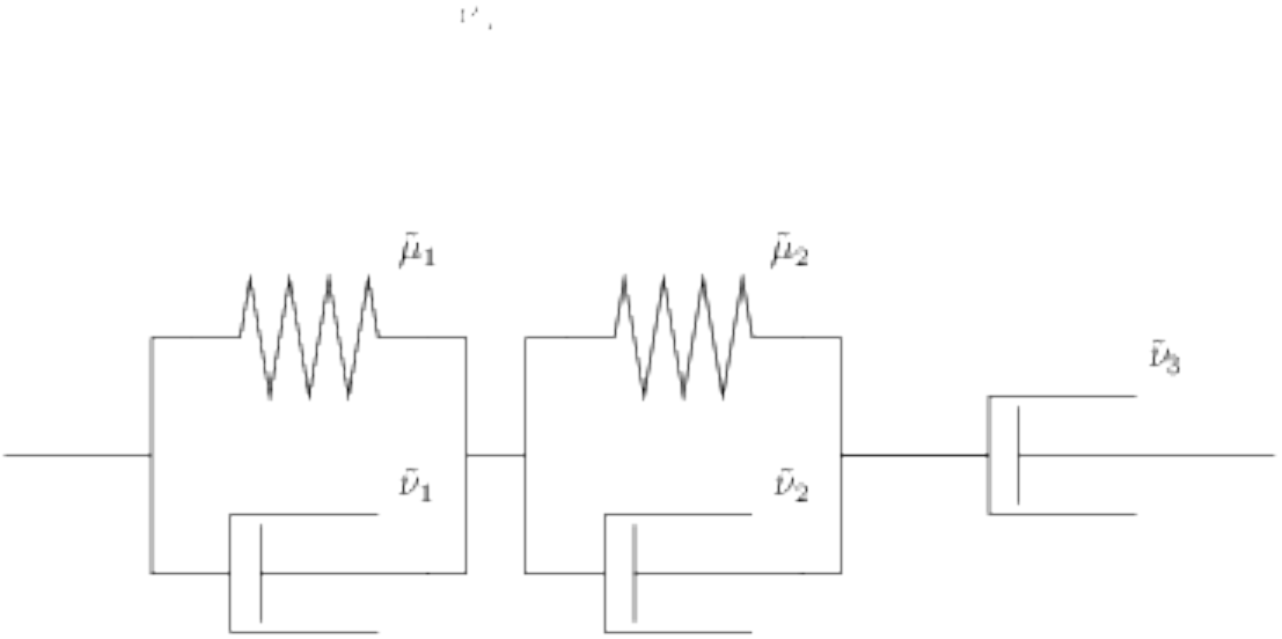}}
 \caption{One dimensional mechanical analogues for a Burgers-type viscoelastic model.}
 \label{fig:burgers-variants}
\end{figure}

The first possible mechanical analogue for Burgers-type model is shown in Figure~\ref{fig:karel}. This mechanical analogue consists of two Maxwell-type elements and an additional dashpot in parallel, see Figure~\ref{fig:karel}. The one-dimensional stress-strain relation for this model has the form
\begin{equation}
  \label{eq:one_karel}
  \sigma + \left(\frac{\nu_1}{\mu_1} + \frac{\nu_2}{\mu_2} \right) \dot \sigma + \frac{\nu_1 \nu_2}{\mu_1 \mu_2} \ddot \sigma
  =
  \left(\nu_1 + \nu_2 + \mu_3 \right) \dot \epsilon 
  +
  \left( \frac{\nu_1 \left(\nu_2+\nu_3\right)}{\mu_1} + \frac{\nu_2 \left(\nu_1+ \nu_3\right)}{\mu_2} \right) \ddot \epsilon + \frac{\nu_1 \nu_2 \nu_3}{\mu_1 \mu_2} \dddot \epsilon
\end{equation}
where $\sigma$ denotes the stress, $\epsilon$ denotes the strain, and the dot denotes the time derivative. Parameters $\nu_1$, $\nu_2$, $\nu_3$ and $\mu_1$, $\mu_2$ are material parameters characterising the elastic moduli and viscosities of the individual springs and dashpots respectively.

For numerical simulations we need to generalise the one-dimensional model into a three dimensional setting. Usually, this step is done as follows. First the Cauchy stress tensor $\cstress$ is for incompressible fluids decomposed to the pressure $p$ and the extra stress tensor $\tensorq{\widehat{\ecstress}}$,
\begin{subequations}
  \label{eq:karel}
\begin{equation}
  \label{eq:1}
  \cstress = -p \tensorq{I} + \tensorq{\widehat{\ecstress}}.
\end{equation}
while the evolution equation for~$\tensorq{\widehat{\ecstress}}$ is obtained from~\eqref{eq:one_karel} by replacing $\sigma$ by $\tensorq{\widehat{\ecstress}}$ and $\dot \epsilon$ by $2 \gradsym$,
 \begin{equation} \label{eq:bugers-karel}
   \tensorq{\widehat{\ecstress}} + \left(\frac{\nu_1}{\mu_1} + \frac{\nu_2}{\mu_2} \right) \fid{\tensorq{\widehat{\ecstress}}} +  \frac{\nu_1 \nu_2}{\mu_1 \mu_2} \fidd{\tensorq{\widehat{\ecstress}}}
   =
   2 \left(\nu_1 + \nu_2 + \mu_3 \right) \gradsym 
   +
   2\left( \frac{\nu_1 \left(\nu_2+\nu_3\right)}{\mu_1} + \frac{\nu_2 \left(\nu_1+ \nu_3\right)}{\mu_2} \right) \fid \gradsym
   +
   2\frac{\nu_1 \nu_2 \nu_3}{\mu_1 \mu_2} \fidd \gradsym,
 \end{equation}
\end{subequations}
where $\gradsym$ denotes the symmetric part of the velocity gradient, $\gradsym=_{\bydefinition}\frac{1}{2} \left (\gradv+\transpose{(\gradv)} \right)$, and the symbol $\fid{\generictensor}$ denotes the upper convected Oldroyd derivative
\begin{equation}
  \label{eq:8}
  \fid{\tensorq{\widehat{\ecstress}}}
  =_{\bydefinition}
  \pd{\tensorq{\widehat{\ecstress}}}{t}
  +
  (\vecv\cdot\nabla)
  \tensorq{\widehat{\ecstress}}
  -
  (\gradv)
  \tensorq{\widehat{\ecstress}}
  -
  \tensorq{\widehat{\ecstress}} \transpose{(\gradv)}
  .
\end{equation}

The second possible mechanical analogue for Burgers-type model is shown in Figure~\ref{fig:paper}. This mechanical analogue consists of two Kelvin--Voigt-type elements and the additional dashpot in series, see Figure~\ref{fig:paper}. The corresponding one-dimensional stress-strain relation reads
\begin{equation} \label{eq:bugers-paper-1d}
  \sigma
  +
  \left(\frac{\tilde\nu_1+\tilde\nu_3}{\tilde \mu_1} + \frac{\tilde\nu_2+\tilde\nu_3}{\tilde \mu_2} \right)
  \dot{\sigma}
  +
  \frac{\tilde\nu_1 \tilde\nu_2 + \tilde\nu_1 \tilde\nu_3 + \tilde\nu_2 \tilde\nu_3}{\tilde \mu_1 \tilde \mu_2}
  \ddot{\sigma}
  = 
   \tilde\nu_3 \dot \epsilon
  +
   \left( \frac{\tilde\nu_1 \tilde\nu_3}{\tilde \mu_1} +  \frac{\tilde\nu_2 \tilde\nu_3}{\tilde \mu_2} \right) \ddot \epsilon
  +
   \frac{\tilde\nu_1 \tilde\nu_2 \tilde\nu_3}{\tilde \mu_1 \tilde \mu_2} \dddot \epsilon,
 \end{equation}
where $\tilde\nu_1$, $\tilde\nu_2$, $\tilde\nu_3$ and $\tilde\mu_1$, $\tilde\mu_2$ again denote material parameters characterising the elastic moduli and viscosities of the individual springs and dashpots respectively. Application of the same approach as before leads to the following three-dimensional generalisation of the one-dimenisonal model,
\begin{equation}
  \label{eq:bugers-paper}
  \tensorq{\widehat{\ecstress}}
  +
  \left(\frac{\tilde\nu_1+\tilde\nu_3}{\tilde \mu_1} + \frac{\tilde\nu_2+\tilde\nu_3}{\tilde \mu_2} \right) \fid{\tensorq{\widehat{\ecstress}}}
  +
  \frac{\tilde\nu_1 \tilde\nu_2 + \tilde\nu_1 \tilde\nu_3 + \tilde\nu_2 \tilde\nu_3}{\tilde \mu_1 \tilde \mu_2} \fidd{\tensorq{\widehat{\ecstress}}}
  = 
  2 \tilde\nu_3 \gradsym
  +
  2 \left( \frac{\tilde\nu_1 \tilde\nu_3}{\tilde \mu_1} +  \frac{\tilde\nu_2 \tilde\nu_3}{\tilde \mu_2} \right) \fid \gradsym
  +
  2 \frac{\tilde\nu_1 \tilde\nu_2 \tilde\nu_3}{\tilde \mu_1 \tilde \mu_2} \fidd \gradsym
  .
\end{equation}

Constitutive relations \eqref{eq:bugers-karel} and \eqref{eq:bugers-paper} have the same form
\begin{equation}
  \label{eq:9}
  \ecstress + c_1 \fid \ecstress + c_2 \fidd \ecstress = c_3 \gradsym + c_4 \fid \gradsym + c_5 \fidd \gradsym
\end{equation}
where $c_i$,  $i=1,\dots,5$ denote material parameters. This observation allows one to compare the different sets of material parameters, that is to express $\nu_1$, $\nu_2$, $\nu_3$ and $\mu_1$, $\mu_2$ in terms of $\tilde\nu_1$, $\tilde\nu_2$, $\tilde\nu_3$ and $\tilde\mu_1$, $\tilde\mu_2$ and \emph{vice versa}. It suffices to solve a system of nonlinear algebraic equations of the type $\frac{\tilde\nu_1+\tilde\nu_3}{\tilde \mu_1} + \frac{\tilde\nu_2+\tilde\nu_3}{\tilde \mu_2} =  \frac{\nu_1}{\mu_1} + \frac{\nu_2}{\mu_2}$ and so forth.

Moreover, the use of the Oldroyd upper convected derivative~\eqref{eq:8}, or for that matter of any objective tensorial derivative, inherently introduces a \emph{nonlinearity} into the constitutive relations, while the original one-dimensional stress-strain is given by a \emph{linear} ordinary differential equation. This is potentially a dangerous step, since \emph{various nonlinear models can reduce to the same linear one-dimensional stress--strain relation}, see for example~\cite{karra.s.rajagopal.kr:development} for a discussion thereof. Consequently, the choice of the right model for the three dimensional computations is not as straightforward as it may seem.

We resolve the issue of the choice of the model as follows. The one-dimensional spring/dashpot model~\eqref{eq:one_karel}, constructed on the basis of the mechanical analogue shown in Figure~\ref{fig:karel}, has been already successfully generalised into the three dimensional setting via a thermodynamically-based procedure introduced by~\cite{rajagopal.kr.srinivasa.ar:thermodynamic} for Maxwell/Oldroyd-B type models. This procedure guarantees the compatibility of the given constitutive relation with the laws of thermodynamics, and it has motivated many further theoretical as well as numerical works concerning viscoelastic rate-type fluids, see for example~\cite{rao.ij.rajagopal.kr:thermodynamic}, \cite{sodhi.js.rao.ij:modeling}, \cite{malek.j.rajagopal.kr.ea:on}, \cite{rehor.m.pusa.v.ea:on}, \cite{kwack.j.masud.a.ea:stabilized} and \cite{hron.j.milos.v.ea:on}. In particular, Burgers-type models constructed via the procedure have been derived and treated numerically in~\cite{tuma.k:identification}, see also~\cite{hron.j.rajagopal.kr.ea:flow}, \cite{malek.j.rajagopal.kr.ea:thermodynamically, malek.j.rajagopal.kr.ea:thermodynamically*1} and~\cite{narayan.spa.little.dn.ea:modelling}. Consequently, in the ongoing numerical simulations, we prefer this well established Burgers-type model to the Burgers-type model constructed from the mechanical analogue shown in Figure~\ref{fig:paper}.

However, the values of the material parameters have been reported by~\cite{sharif-kashani.p.hubschman.j.ea:rheology} for the \emph{one-dimensional} model~\eqref{eq:bugers-paper-1d}, see Table~\ref{tab:parameters-original}. (\cite{sharif-kashani.p.hubschman.j.ea:rheology} did not write down a fully three dimensional model.) Therefore, if we want to use the model based on the equivalent one-dimensional model~\eqref{eq:one_karel}, we have to first convert the set of parameters $\tilde\nu_1$, $\tilde\nu_2$, $\tilde\nu_3$ and $\tilde\mu_1$, $\tilde\mu_2$ to the set $\nu_1$, $\nu_2$, $\nu_3$ and $\mu_1$, $\mu_2$ via the procedure described above. This gives the parameter values reported in Table~\ref{tab:parameters}. In both tables the symbols $\tau_i$ and $\tilde{\tau}_i$ denote the relaxation time, that is the viscosity divided by the elastic shear modulus, $\tau_i =_{\bydefinition} \frac{\nu_i}{\mu_i}$, $i=1,2$ and $\tilde{\tau}_i =_{\bydefinition} \frac{\tilde{\nu}_i}{\tilde{\mu}_i}$, $i=1,2$.

\input{table-1}
\input{table-2}

Finally, we note that in~\eqref{eq:karel} it is convenient to introduce a decomposition
\begin{equation}
  \label{eq:10}
  \tensorq{\widehat{\ecstress}} = 2\nu_3\gradsym + \tensorq{S},
\end{equation}
that allows one to rewrite~\eqref{eq:karel} in an equivalent form as
\begin{subequations}
  \label{eq:11}
  \begin{align}
    \label{Burgers_model1a}
    \cstress
    &=
      -p \identity + 2\nu_3\gradsym + \tensorq{S}, \\
    \label{Burgers_model1b}
    \fidd{\tensorq{S}}
    +
    \left(\frac{\mu_1}{\nu_1}+\frac{\mu_2}{\nu_2}\right) \fid{\tensorq{S}}
    +
    \frac{\mu_1 \mu_2}{\nu_1 \nu_2}\tensorq{S}
    &=
      2\left(\frac{\mu_1 \mu_2}{\nu_2}+\frac{\mu_1 \mu_2}{\nu_1}\right)\gradsym
      +
      2(\mu_1+\mu_2)\fid{\gradsym}.
  \end{align}
\end{subequations}
Using this reformulation of~\eqref{eq:karel}, it is straightforward to see that the model indeed provides a generalisation of the standard Navier--Stokes fluid model where $\cstress = -p \identity + 2\nu_3\gradsym$.

\subsection{Sclera and lens}
\label{sec:sclera-lens}
The sclera and the lens are modelled as hyperelastic solids, see~\cite{truesdell.c.noll.w:non-linear*1}, with the strain energy density $W$. Following \citet{grytz.r.fazio.ma.ea:material} the response of human sclera is described by the strain energy density in the form
\begin{equation}
  \label{eq:12}
  W = \frac{1}{2} \mu\left(J^{-2/3}\Tr\rcg-3\right) + \frac{1}{2} \kappa(\ln J)^2,
\end{equation}
where $\fgrad$ denotes the deformation gradient, $J = _{\bydefinition} \det \fgrad$, and $\rcg=_{\bydefinition}\fgrad^{\rm T}\fgrad$. Parameter $\mu$ is referred to as the elastic shear modulus, while $\kappa$ is referred to as the elastic bulk modulus. Since both sclera and lens are almost incompressible, we enforce the incompressibility by taking $\kappa=1000\,\mu$ which consequently makes $J\doteq 1$, and thus $\rho\doteq\rho_0=const.$ Parameter values for human sclera and lens are shown in Table~\ref{tab:parameters}.


%% file: table-1.tex
\begin{table}[h] 
\begin{center}
  \begin{tabular}[t]{llcr@{$\:\times\:$}l }
    \toprule 
    Parameter & Description & Units & \multicolumn{2}{c}{Value} \\
    \midrule
    $\tilde{\nu}_3$ & viscosity & $\unit{Pa} \cdot \unit{s}$ &  $1.057$ & $10^{3}$ \\ 
    $\tilde{\mu}_1$ & first elastic shear modulus & $\unit{Pa}$ & $1.66$ & $10^{0}$  \\
    $\tilde{\mu}_2$ & second elastic shear modulus & $\unit{Pa}$ & $1.14$ & $10^{0}$  \\
    $\tilde{\tau}_1$ & first relaxation time & $\unit{s}$ &  $1.97$ & $10^{0}$ \\ 
    $\tilde{\tau}_2$ & second relaxation time & $\unit{s}$ &  $90$ & $10^{0}$  \\
    \bottomrule
  \end{tabular}
  \medskip
  \caption{Material parameter values for the vitreous humour as reported in~\cite{sharif-kashani.p.hubschman.j.ea:rheology}, one-dimensional model~\eqref{eq:bugers-paper-1d}.}
  \label{tab:parameters-original}
\end{center}
\end{table}


%% file: table-2.tex
\begin{table}[h] 
\begin{center}
  \begin{tabular}[t]{llcr@{$\:\times\:$}l cr@{$\:\times\:$}l}
    \toprule 
    Parameter & Description & Units & \multicolumn{2}{c}{Value} & Source &  \multicolumn{2}{c}{Computation} \\
    \midrule
    $\rho$ & density & $\unitfrac{kg}{m^3}$ & $1.0053 - 1.0089$ & $10^{3}$ & \cite{black2013handbook} & $1.007$ & $10^{3}$\\
        $\nu_3$ & viscosity & $\unit{Pa} \cdot \unit{s}$ &  $2.37$ & $10^{0}$ & \cite{sharif-kashani.p.hubschman.j.ea:rheology} & $2.37$ & $10^{0}$\\
        $\mu_1$ & shear modulus & $\unit{Pa}$ & $6.45$ & $10^{-1}$ & \cite{sharif-kashani.p.hubschman.j.ea:rheology} & $6.45$ & $10^{-1}$\\
        $\mu_2$ & shear modulus & $\unit{Pa}$ & $8.98$ & $10^{-1}$  & \cite{sharif-kashani.p.hubschman.j.ea:rheology} & $8.98$ & $10^{-1}$\\
        $\tau_1$ & relaxation time & $\unit{s}$ &  $1.60$ & $10^{3}$  & \cite{sharif-kashani.p.hubschman.j.ea:rheology} & $1.60$ & $10^{3}$\\
        $\tau_2$ & relaxation time & $\unit{s}$ &  $25.06$ & $10^{0}$  & \cite{sharif-kashani.p.hubschman.j.ea:rheology} & $25.06$ & $10^{0}$\\
    \midrule
    $\mu$ & shear modulus, sclera & $\unit{Pa}$ &  $330$ & $10^{3}$  & \cite{grytz.r.fazio.ma.ea:material} & $330$ & $10^{3}$\\
        $\mu$ & shear modulus, lens & $\unit{Pa}$ & $0.19 - 59.6$ & $10^{3}$ & \cite{wilde.gs.burd.hj.ea:shear} & $10$ & $10^{3}$\\
        $\rho$ & density, sclera & $\unitfrac{kg}{m^3}$ & $1.076$ & $10^{3}$ & \cite{dens} & $1.05$ & $10^{3}$\\
        $\rho$ & density, lens & $\unitfrac{kg}{m^3}$ & $1.104$ & $10^{3}$ & \cite{dens} & $1.05$ & $10^{3}$\\
    \bottomrule
  \end{tabular}
  \medskip
  \caption{Material parameter values for the vitreous humour, the sclera and the lens. Material parameters for Burgers-type viscoelastic model~\eqref{eq:bugers-karel} has been calculated from the experimental data by~\cite{sharif-kashani.p.hubschman.j.ea:rheology}, see Table~\ref{tab:parameters-original}.  Symbols $\tau_i$ and $\tilde{\tau}_i$ denote the relaxation time, that is the viscosity divided by the elastic shear modulus, $\tau_i =_{\bydefinition} \frac{\nu_i}{\mu_i}$, $i=1,2$ and $\tilde{\tau}_i =_{\bydefinition} \frac{\tilde{\nu}_i}{\tilde{\mu}_i}$, $i=1,2$. Column ``Computation'' shows the material parameter values used in the reported numerical simulation.}
  \label{tab:parameters}
\end{center}
\end{table}


%% file: governing.tex
\renewcommand{\displacementu}{\vec{u}}
\newcommand{\pdd}[2]{\ensuremath{\frac{\partial^2 {#1}}{\partial {#2}^2}}}
\newcommand{\fpkstress}{\tensorq{P}}

\subsection{Governing equations for the flow of vitreous humour}
Vitreous humour is described by the three-dimensional Burgers-type model~\eqref{eq:11}. Since this is a fluid-like model, it is convenient to formulate the governing equations in the Eulerian description. The balance of mass and the balance of linear momentum take the form
\begin{subequations}
  \label{eq:14}
  \begin{align}
    \divergence\vecv&=0,\label{divv0}\\
    \rho\left(\pd{\vecv}{t}+(\vecv\cdot\nabla)\vecv\right)&=\divergence\cstress,\label{bal_linmom}
  \end{align}
\end{subequations}
where $\vecv$ denotes the velocity, and where we have already exploited the fact that we are dealing with an incompressible homogenous fluid. The Cauchy stress tensor is given by~\eqref{eq:11}.

However, formulation~\eqref{eq:11} is not convenient for numerical treatment. It turns out, see~\cite{tuma.k:identification} and~\cite{hron.j.rajagopal.kr.ea:flow}, that~\eqref{eq:11} can be rewritten as
\begin{subequations}
  \label{eq:13}
  \begin{equation}
    \cstress=-p\identity+2\nu_3\gradsym+\mu_1(\lcg_1-\identity)+\mu_2(\lcg_2-\identity),\label{Burgers_model2a}
  \end{equation}
  where the left Cauchy--Green tensors $\lcg_1$ and $\lcg_2$ satisfy 
  \begin{align}
    \fid{{\lcg}_1} + \frac{\mu_1}{\nu_1} \left( \lcg_1 - \identity \right) &= \tensorq{0},\label{Burgers_model2b}\\
    \fid{{\lcg}_2} + \frac{\mu_2}{\nu_2} \left( \lcg_2 - \identity \right) &= \tensorq{0}\label{Burgers_model2c}.
  \end{align}
\end{subequations}
This form is convenient for several reasons. First, the \emph{second} order differential equation~\eqref{Burgers_model1b} for $\ecstress$ has been replaced by two \emph{first} order differential equations for $\lcg_1$ and $\lcg_2$, which simplifies time stepping algorithm. Second, quantities $\lcg_1$ and $\lcg_2$ correspond to the ``state'' of the springs in Figure~\ref{fig:karel}, which means that it is \emph{easy to specify the initial conditions} for $\lcg_1$ and $\lcg_2$. Indeed, in the (initial) undeformed state, one has to set $\lcg_1 = \identity$ and $\lcg_2 = \identity$, see~\cite{hron.j.rajagopal.kr.ea:flow} for details.

\subsection{Governing equations for the deformation of the lens and the sclera}
The governing equations for the lens and sclera are formulated in the Lagrangian description. The balance of mass and linear momentum take the the form
\begin{subequations}
  \label{eq:15}
\begin{align}
  \rho_0&=J\rho, \label{solid_model1}\\
  \rho_0\pdd{\displacementu}{t}&=\Divergence\fpkstress,\label{solid_model2}
\end{align}
where $\rho$ denotes the density in the current configuration, $\rho_0$ denotes the density in the reference configuration, and $\displacementu$ is the displacement, and $\fgrad = \identity+\nabla\displacementu$ denotes the deformation gradient. The symbol $\fpkstress$ stands for the first Piola--Kirchhoff stress tensor, which is related to the strain energy density $W$ via the formula
\begin{equation}
  \fpkstress =\pd{W}{\fgrad}\label{FPKstress_def},
\end{equation}
see for example~\cite{truesdell.c.noll.w:non-linear*1}, where the specific strain energy density $W$ for the sclera/lens is given by~\eqref{eq:12}.
\end{subequations}


%% file: numerics.tex
The overall mechanical response of the eye stems from an interaction of the flowing vitreous humour and the deforming sclera and the lens. In our approach we aim to compute the whole problem on a fixed mesh.

The governing equations for the solid parts~\eqref{eq:15} are formulated in the Lagrangian description, and are solved for in the fixed reference domain $\Omega_X^{1}$ and $\Omega_X^{2}$ for the unknown displacement $\displacementu$ and velocity field $\vec{v}$. (Symbol $\Omega_X^1$ denotes the reference domain occupied by the lens, while $\Omega_X^2$ denotes the reference domain occupied by the sclera, see Figure~\ref{fig:geometry-a}.) On the other hand, the governing equations for the vitreous humour~\eqref{eq:14} are formulated in the Eulerian description, that is they hold in the domain that is at the given instant occupied by the fluid, and they must be solved for the unknowns $\vecv$, $p$, $\lcg_1$, $\lcg_2$. However the domain occupied by the fluid is changing in time. The problem of moving domain is addressed by the the arbitrary Lagrangian-Eulerian (ALE) method, see for example~\cite{donea.j.huerta.a.ea:arbitrary} and~\cite{scovazzi.g.hughes.t:lecture}, that is based on the mapping of the changing domain (moving mesh) to the fixed domain $\Omega_{\chi}$. This introduces an additional unknown -- the deformation of the mesh $\hat{\displacementu}$.

Using the ALE method, the fluid-structure interaction is computed using a monolithic approach, where all unknowns fields for the solid and fluid part are solved for simultaneously. This monolithic approach has been successfully applied in the case of interaction between a neo-Hookean solid and a Newtonian fluid in a two-dimensional setting, see for example~\cite{razzaq.m.hron.j.ea:numerical}, as well as in a three dimensional setting, see for example~\cite{hron.j.madlik.m:fluid-structure}.

\subsection{Viscoelastic part}
The governing equations for the viscoelastic fluid are mapped by ALE method to a fixed domain $\Omega_\chi$. In a two-dimensional setting,  the Burgers-type model has been already treated by the ALE method, see~\cite{hron.j.rajagopal.kr.ea:flow}. In contrast to~\cite{hron.j.rajagopal.kr.ea:flow}, here we deal with a \emph{fully three dimensional fluid-structure interaction problem}, and we need to take into account the interaction with two solid materials. The ALE method leads to the following weak formulation that is used for finite element discretisation of the corresponding equations
\begin{subequations}
  \label{eq:16}
\begin{align}
\int_{\Omega_\chi}\hat{J}\Tr\left((\nabla_\chi\vecv)\hat{\fgrad}^{-1}\right) q\ \diff\chi&=0,\\
\int_{\Omega_\chi}\hat{J}\rho \left[\pd{\vecv}{t}+(\nabla_\chi\vecv)\left(\hat{\fgrad}^{-1}\left(\vecv-\pd{\hat{\displacementu}}{t}\right)\right)\right]\cdot\vec{q}\ \diff\chi+\int_{\Omega_\chi}\hat{J}\hat{\cstress}\hat{\fgrad}^{-\rm T}\cdot\nabla_\chi\vec{q}\ \diff\chi&=0,\\
\hat{\cstress}=-p\identity+\nu_3\left((\nabla_\chi\vecv)\hat{\fgrad}^{-1}+\hat{\fgrad}^{-\rm T}(\nabla_\chi\vecv)^{\rm T}\right)+\mu_1(\lcg_1-\identity)+\mu_2(\lcg_2&-\identity),\\
\int_{\Omega_\chi} \hat{J}\left[\pd{\lcg_1}{t}+(\nabla_\chi\lcg_1)\left(\hat{\fgrad}^{-1}\left(\vecv-\pd{\displacementu}{t}\right)\right)-(\nabla_\chi\vecv)\hat{\fgrad}^{-1}\lcg_1-\lcg_1\hat{\fgrad}^{-\rm T}(\nabla_\chi\vecv)^{\rm T}+\frac{\mu_1}{\nu_1}(\lcg_1-\identity)\right]\cdot\tensorq{Q}_1\ \diff\chi&=0,\\
\int_{\Omega_\chi} \hat{J}\left[\pd{\lcg_2}{t}+(\nabla_\chi\lcg_2)\left(\hat{\fgrad}^{-1}\left(\vecv-\pd{\displacementu}{t}\right)\right)-(\nabla_\chi\vecv)\hat{\fgrad}^{-1}\lcg_2-\lcg_2\hat{\fgrad}^{-\rm T}(\nabla_\chi\vecv)^{\rm T}+\frac{\mu_2}{\nu_2}(\lcg_2-\identity)\right]\cdot\tensorq{Q}_2\ \diff\chi&=0,\\
\int_{\Omega_\chi} \nabla_\chi\hat{\displacementu}\cdot\nabla_\chi\hat{\vec{t}}\ \diff\chi&=0.
\end{align}
\end{subequations}
where $q,\vec{q},\tensorq{Q}_1,\tensorq{Q}_2, \vec{t}$ are the admissible test functions, $\hat{\displacementu}$ is the corresponding ALE displacement, $\hat\fgrad$ its  deformation gradient $\hat{\fgrad}=\identity+\nabla_\chi\hat{\displacementu}$, $\hat{J}=\det{\hat{\fgrad}}$ and $\nabla_\chi$ denotes the gradient operator in the ALE frame. As it is apparent from the last equation, the mesh motion is governed by the equation
\begin{equation}
  \label{eq:17}
  -\Delta_\chi\hat{\displacementu} = \vec{0},
\end{equation}
which is the standard choice in the ALE method. All points on the boundary $\partial\Omega_\chi$ are material points, thus it holds $\pd{\hat{\displacementu}}{t}=\vecv$.
This relation is enforced by employing the method of Lagrange multipliers that is based on finding the extreme points of the Lagrange function
\begin{equation}
  \label{eq:19}
L(\vec{\lambda},\hat{\displacementu},\vecv)=_{\bydefinition}\left(\pd{\hat{\displacementu}}{t}-\vecv\right)\cdot\vec{\lambda}
\end{equation}
where $\vec{\lambda}$ denotes the vector of Lagrange multipliers that are defined on $\partial\Omega_\chi$ only which does not increase a lot the size of problem. Note that~\eqref{eq:17} implies that $\partial\Omega_\chi$ is composed of material points, and thus the transformed Cauchy stress tensor $\hat{J}\hat{\cstress}\hat{\fgrad}^{-\rm T}$ is, on the boundary $\partial\Omega_\chi$, the first Piola-Kirchhoff stress tensor.
 
\subsection{Solid part}
The governing equations for the solid parts are solved in $\Omega_X^1$ and $\Omega_X^2$, and they take the form
\begin{subequations}
  \begin{align}
    \int_{\Omega_X^{i}}\left(\pd{\displacementu^i}{t}-\vecv^i\right)\cdot\vec{q}_1\ \diff X&=0,\\
    \int_{\Omega_X^{i}}\rho_0^{i}\pd{\vecv^i}{t}\cdot\vec{q}_2\ \diff X+\int_{\Omega_X^{i}}\fpkstress^i\cdot\nabla_X\vec{q}_2\ \diff X&=0,
  \end{align}
\end{subequations}
where $i=1,2$, and where $\vec{q}_1,\vec{q}_2$ are admissible test functions, $\displacementu^i$, $\vec{v}^i$ and $\fpkstress^i$ denote the displacement/velocity/first Piola--Kirchhoff stress in the domain occupied by the sclera and the lens respectively, and $\rho_0^i$ denotes the reference density of the sclera and the lens respectively.

\subsection{Interaction between the vitreous body, the sclera and the lens}
The interaction between the solids occupying the domains $\Omega_X^1$ and $\Omega_X^2$ and the viscoelastic fluid occupying the domain $\Omega_\chi$ takes place on the interfaces between the domains. On the interfaces we enforce the continuity of the displacements and the stresses, that is
\begin{subequations}
  \label{eq:18}
  \begin{align}
    \displacementu_i&=\hat{\displacementu} \hspace*{-5mm} & &\text{and} & \hspace*{-5mm} \fpkstress_i\vec{N}&=\hat{J}\cstress\hat{\fgrad}^{-\rm T}\vec{N}\quad\text{on}\ \partial\Omega_\chi\cap\partial\Omega_X^{i}\ (\text{interface\ between\ fluid\ and\ solid}),\quad i=1,2,\\
    \displacementu_1&=\displacementu_2 \hspace*{-5mm} & &\text{and} & \hspace*{-5mm} \fpkstress_1\vec{N}&=\fpkstress_2\vec{N}\quad\text{on}\ \partial\Omega_X^1\cap\partial\Omega_X^2\ (\text{interface\ between\ two\ solids}),
  \end{align}
\end{subequations}
where $\vec{N}$ is the unit normal vector to the interface. 

\subsection{Implementation}
Weak forms of the governing equations~\eqref{eq:14} and~\eqref{eq:15} are discretised in space using the finite element method, while the time derivatives are approximated with the backward Euler method. The three-dimensional domains $\Omega_\chi$, $\Omega_X^1$, $\Omega_X^2$ are approximated by $\Omega_{\chi, h}$, $\Omega_{X,h}^1$, $\Omega_{X,h}^2$ and discretised by regular hexahedra, see Figure~\ref{fig:mesh}. The velocity $\vecv$ is approximated by $H_1$ elements, other unknowns $\lcg_1$, $\lcg_2$, $\displacementu$ and $\hat{\displacementu}$  are also approximated by $H_1$ elements, while the pressure $p$ is approximated by piecewise constant elements $P_0$.

The finite element method is implemented in the AceGen/AceFEM system~\cite{korelc.j:Multi,korelc.j:Automation}. AceGen is a code generation system, and AceFEM is a finite element environment that uses the generated code. The provided automatic differentiation feature is used for the computation of the first Piola--Kirchhoff stress tensor~\eqref{FPKstress_def}, and the exact tangent matrix used by the Newton solver to treat the non-linearities. The consequent set of linear equations is solved by conjugate gradient squared iterative solver, see~\cite{sonneveld.p:CGS}, that is a part of the package Intel MKL Pardiso. The linear solver finishes when the relative residuum of the linear system reaches $10^{-4}$ and the stopping criterion for the Newton solver is $10^{-9}$.


%% file: results.tex
The numerical code introduced above has been used in solving the boundary value problem described in Section~\ref{sec:problem-description}. In order to identify the impact of the choice of a rheological model for the vitreous humour on the response of the eye, we have solved the boundary value problem for two rheological models.

In the first numerical experiment, the vitreous humour has been described by the simple \emph{Navier--Stokes fluid} model, while in the other numerical experiment we have used the \emph{Burgers-type model} introduced in Section~\ref{sec:vitreous-humour}. (Note that the Navier--Stokes model corresponds to the Burgers-type model with $\mu_1=0$ and $\mu_2=0$, hence the computation for the Navier--Stokes model is just a special case of the computation for the Burgers-type model.) The comparison of these two rheological models is of interest from the practical point of view, since the Navier--Stokes rheological model corresponds to a pathological vitreous humour, while the Burgers-type model describes a healthy vitreous humour. Using the different rheological models for the vitreous humour, we have compared the quantitative and qualitative characteristics of the obtained numerical solutions.

\begin{figure}[h]
\begin{center}
\includegraphics[width=0.4\textwidth]{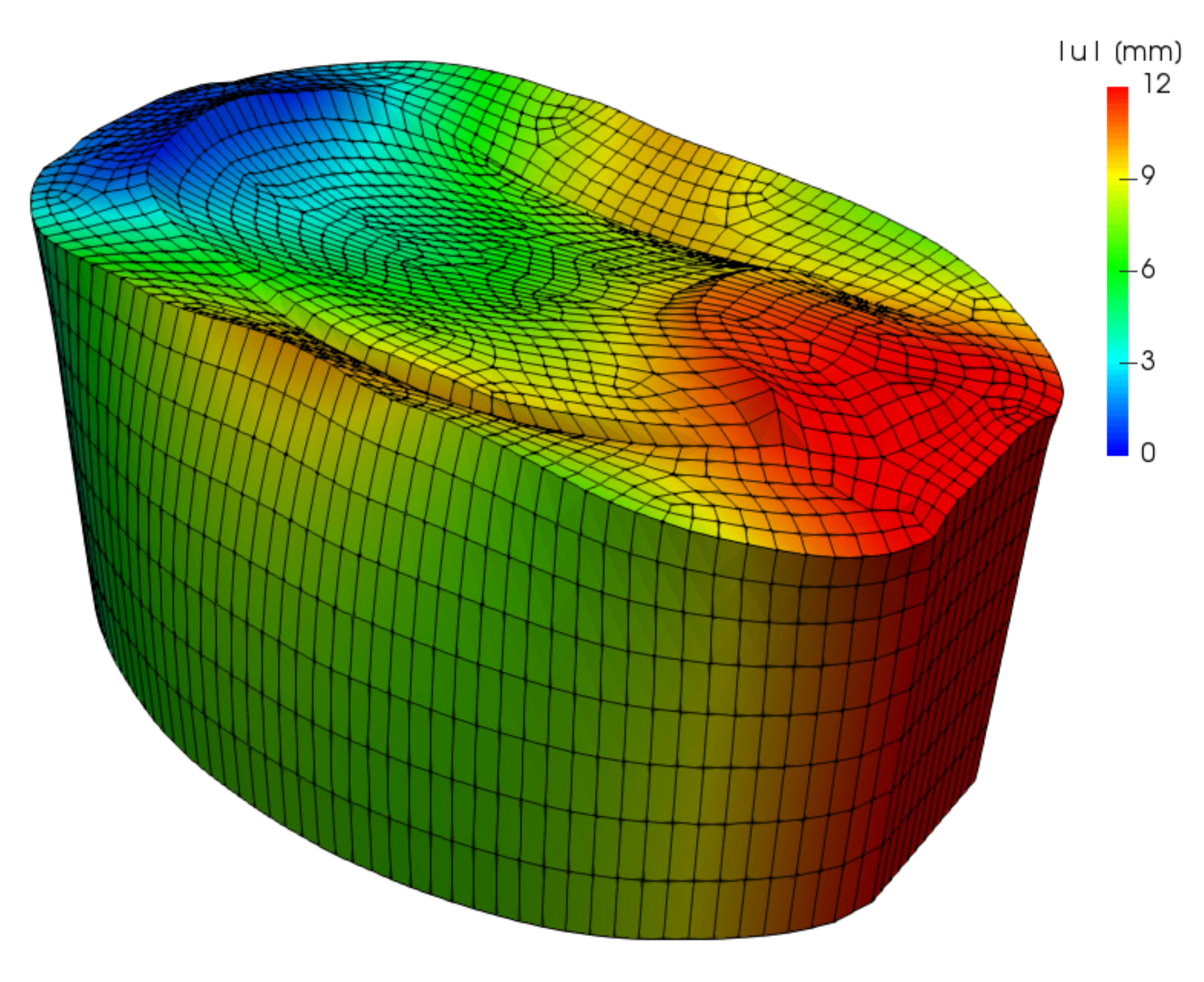}
\end{center}
\caption{Solution at $t=\unit[480]{s}$, displacement magnitude mapped to the current configuration.}
\label{fig:displacement}
\end{figure}

\begin{figure}[h]
  \begin{center}
    \subfloat[\label{fig:response-plots-force} Force $F_x$, see~\eqref{eq:20}, that is necessary to maintain the prescribed displacement. (Curves for the Navier--Stokes fluid and the Burgers-type fluid overlap.)]{\includegraphics[width=0.45\textwidth]{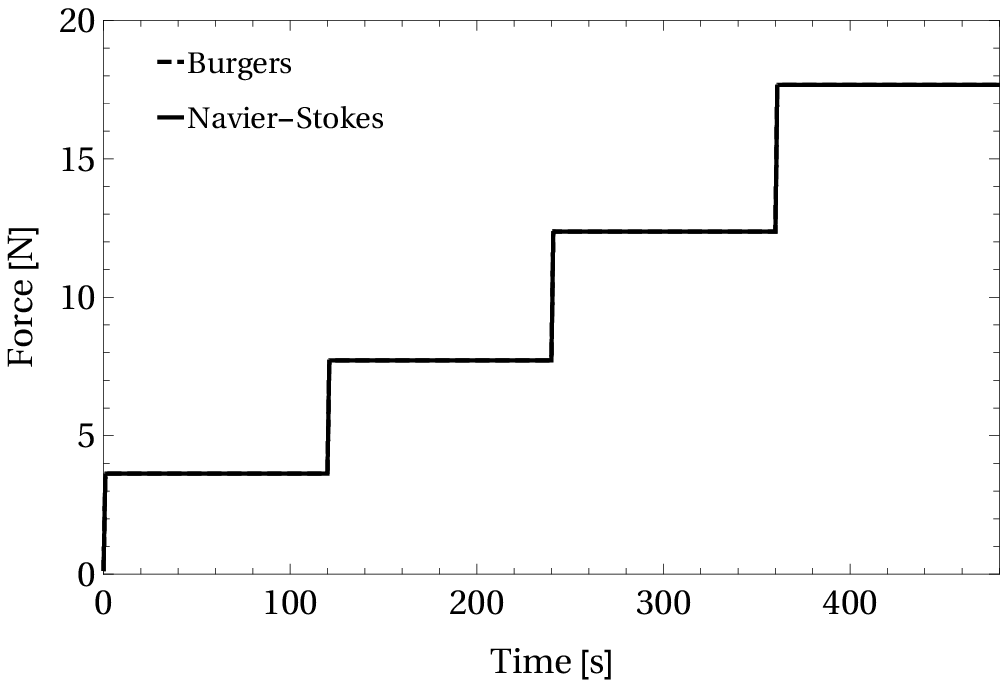}}
    \qquad
    \subfloat[\label{fig:response-plots-displacement} Displacement $u_z$ of the test particle at top in the middle of the vitreous.]{\includegraphics[width=0.45\textwidth]{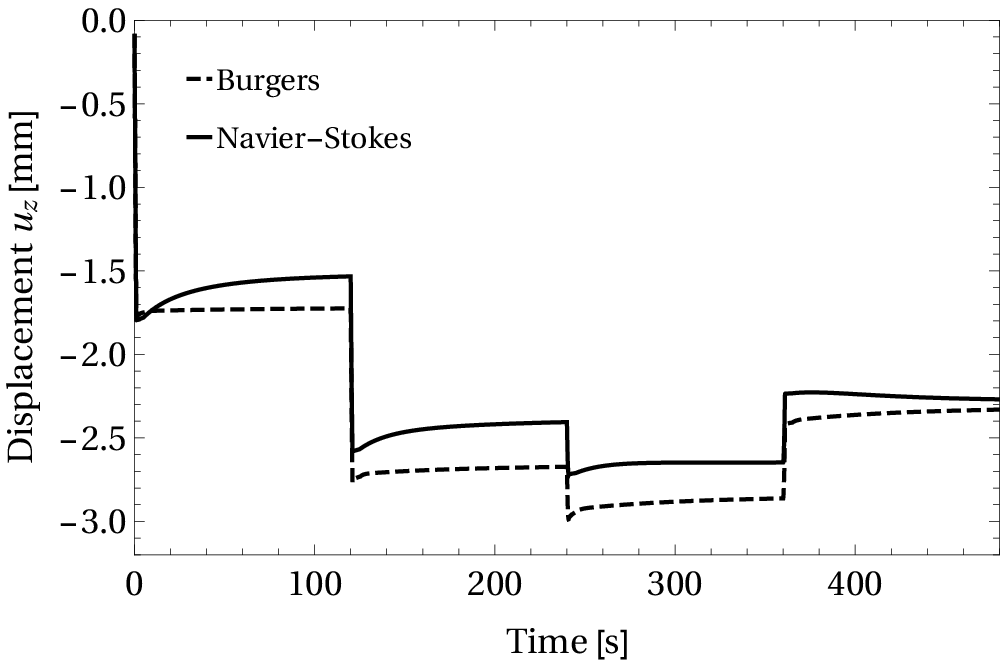}}
  \end{center}
  \caption{Response to the prescribed displacement of the boundary $\Gamma_{\mathrm{contact}_1}$, see Figure~\ref{fig: Prescribed deformation}. Vitreous humour modelled either as the Navier--Stokes fluid of the Burgers-type viscoelastic fluid.}
  \label{fig:response-plots}
\end{figure}

First, given the displacement of the boundary $\Gamma_{\mathrm{contact}_1}$, see Figure~\ref{fig: Prescribed deformation} and Section~\ref{sec:problem-description} for details, we have been interested in the overall force response and the overall displacement induced inside the eye. In particular, using the computed solution we have computed the force that is necessary to maintain the prescribed displacement of the boundary, that is 
\begin{equation}
  \label{eq:20}
  F_x = \vectordot{\left( - \int_{\Gamma_{\mathrm{contact}_1}} \fpkstress \vec{N} \surfacees \right)}{\bvecx},
\end{equation}
and the displacement $u_z$ of a fictitious test particle placed at the top and in the middle of the vitreous. The overall displacement field $\displacementu$ at the end of the test is shown in Figure~\ref{fig:displacement}. The plots of $F_x$ and $u_z$ are shown in Figure~\ref{fig:response-plots}. We see that the overall force response is almost independent on the choice of the rheological model. This is an expected result, since the force response must be clearly dominated by the elastic response of the sclera and the lens. The difference between the two rheological models starts to be apparent at the level of displacement field in the vitreous humour, see Figure~\ref{fig:response-plots-displacement}, although the difference is relatively small.

\begin{figure}[h]
\begin{center}
\includegraphics[width=0.6\textwidth]{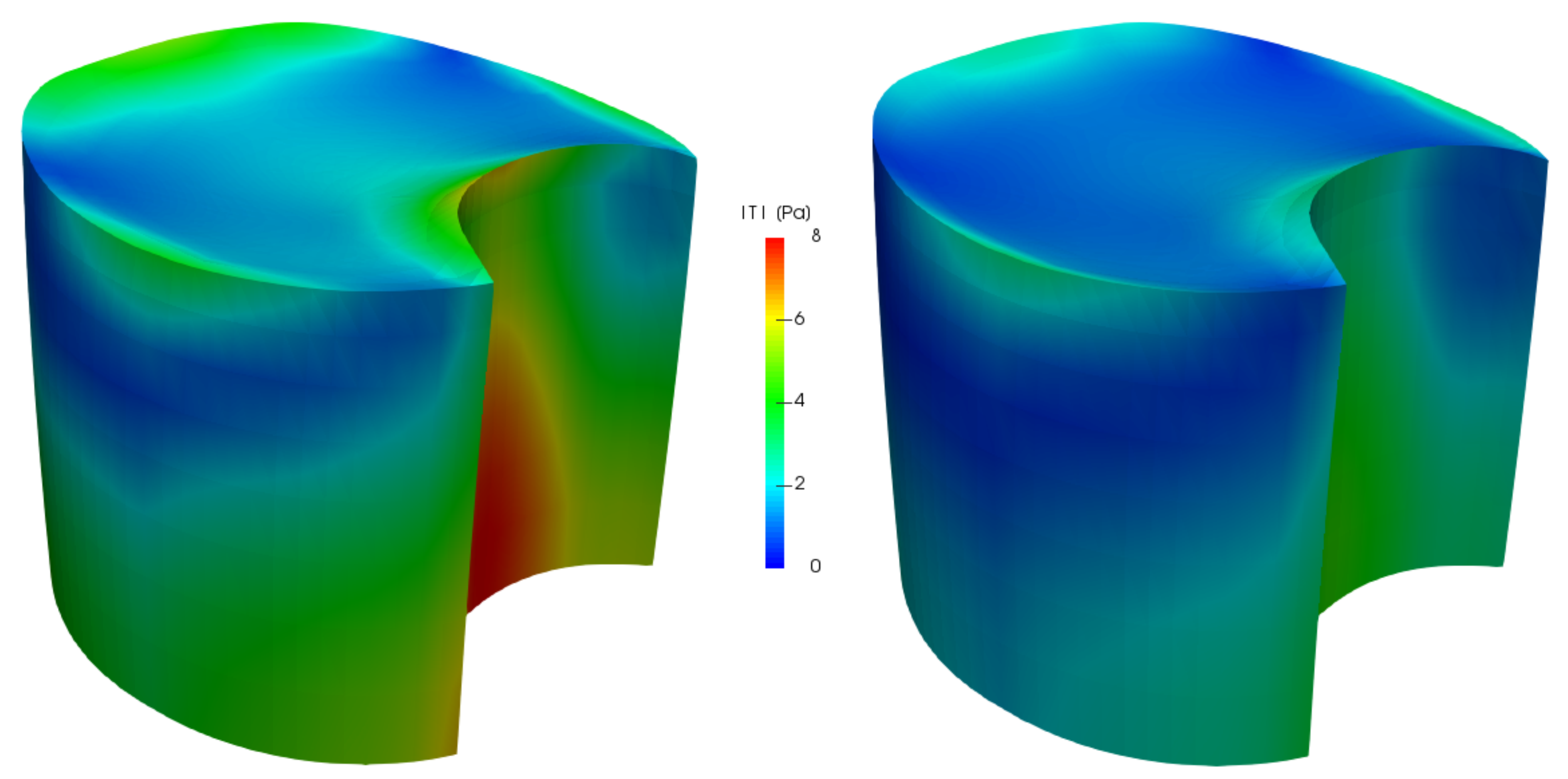}
\end{center}
\caption{Magnitude of the Cauchy stress tensor $\cstress$ at $t=\unit[120.5]{s}$ (half of the second prolongation), current configuration. Vitreous humour modelled using the Burgers-type viscoelastic model (left) and the standard Navier--Stokes model (right).}
\label{fig:stress-space}
\end{figure}

\begin{figure}[h]
\begin{center}
\subfloat[Cross-section with the plane $z=0$.]{\includegraphics[width=0.8\textwidth]{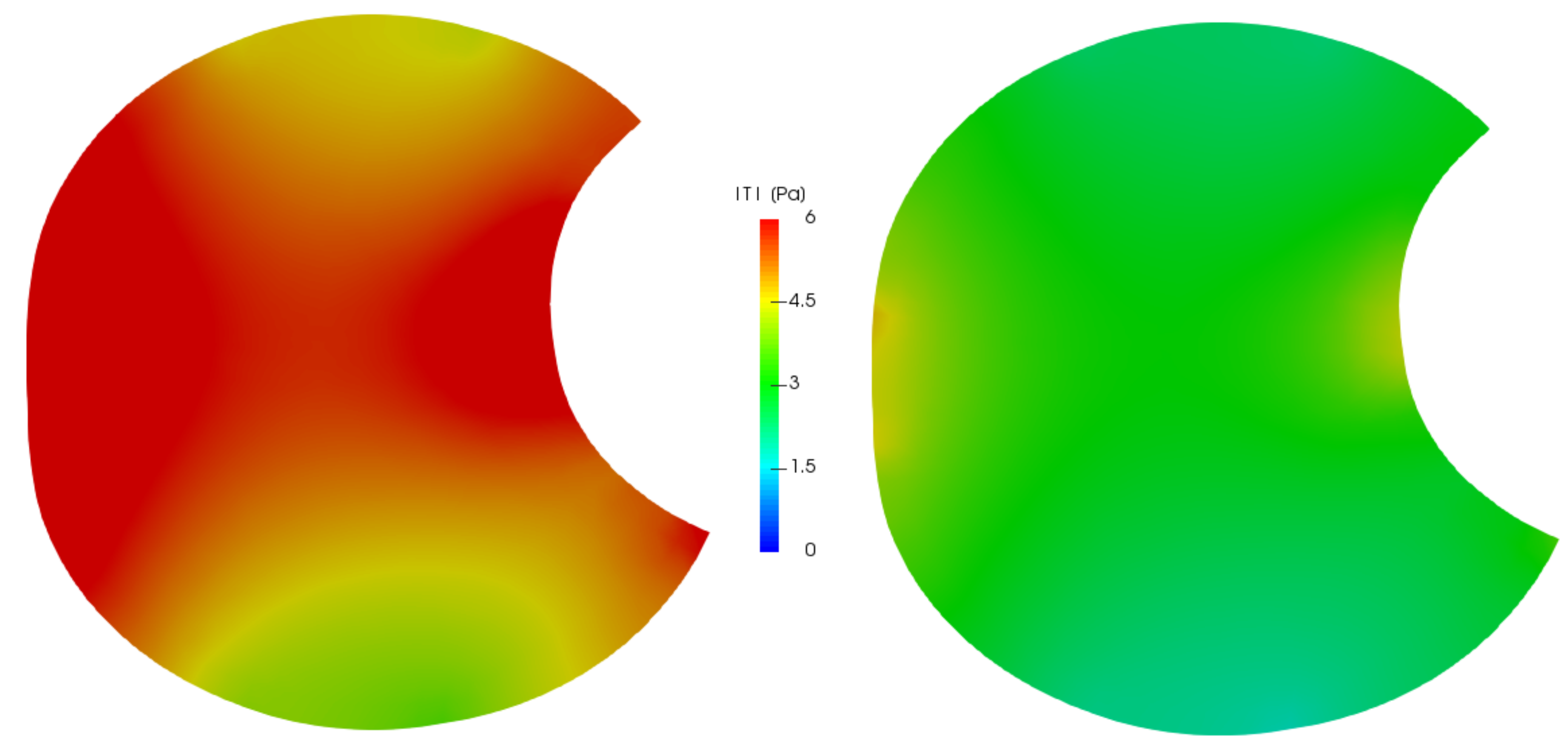}}\\
\subfloat[Cross-section with the plane $z=\frac{h}{2}$.]{\includegraphics[width=0.8\textwidth]{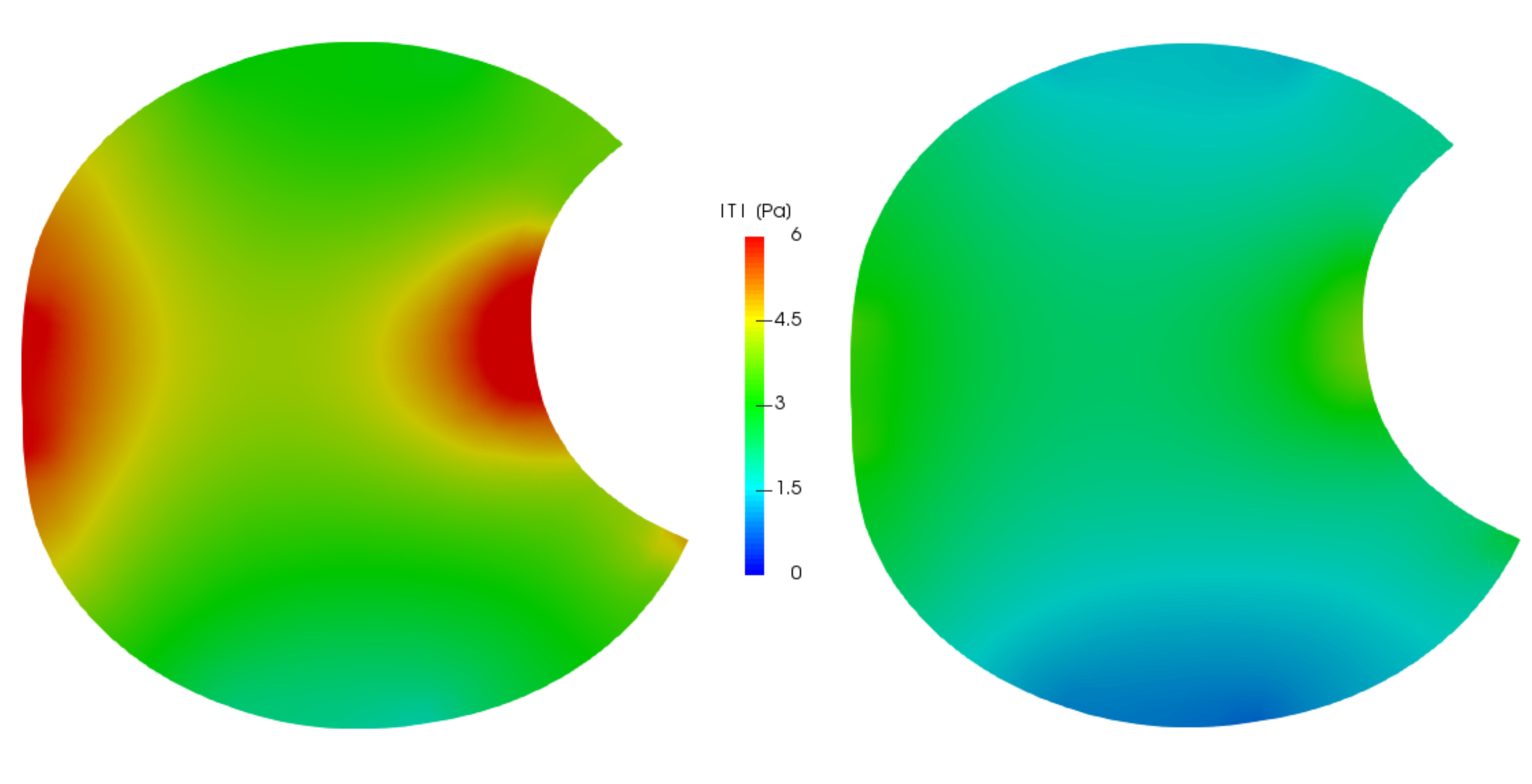}} 
\end{center}
\caption{
  Magnitude of the Cauchy stress tensor $\cstress$ at $t=\unit[120.5]{s}$ (half of the second prolongation), current configuration, cross-section with the plane $z=0$ and $z=\frac{h}{2}$. Vitreous humour modelled using the Burgers-type viscoelastic model (left) and the standard Navier--Stokes model (right).}
\label{fig:stress-cross-section}
\end{figure}

Second, given the displacement of the boundary $\Gamma_{\mathrm{contact}_1}$, see Figure~\ref{fig: Prescribed deformation} and Section~\ref{sec:problem-description} for details, we have been interested in the stress distribution in the vitreous humour. The magnitude of the Cauchy stress tensor $\cstress$ in the vitreous humour is shown in Figure~\ref{fig:stress-space}. The figure shows that the stress distribution in the vitreous humour is significantly influenced by the choice of the rheological model for the vitreous humour, while the deformation of the domain occupied by the vitreous humour is almost the same. This is clearly apparent from the additional plots showing the stress distribution at given cross-sections of the domain, see Figure~\ref{fig:stress-cross-section}. We can notice that the magnitude of the Cauchy stress tensor can be as much as two times higher if one compares the results predicted by the Navier--Stokes model and the results predicted by the Burgers-type model.

\begin{figure}[h]
\begin{center}
\subfloat[Cross-section with the plane $z=0$.]{\includegraphics[width=0.8\textwidth]{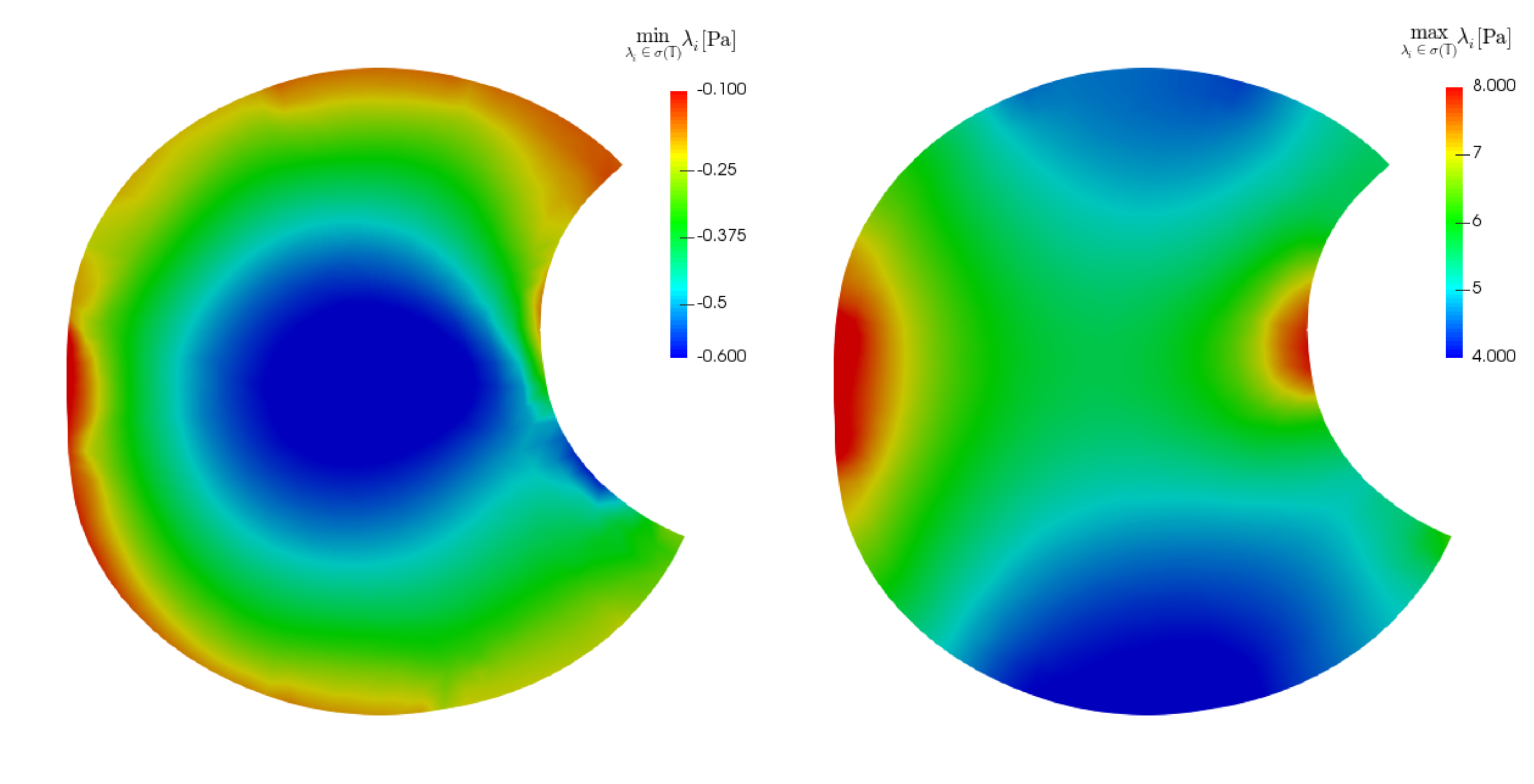}}\\
\subfloat[Cross-section with the plane $z=\frac{h}{2}$.]{\includegraphics[width=0.8\textwidth]{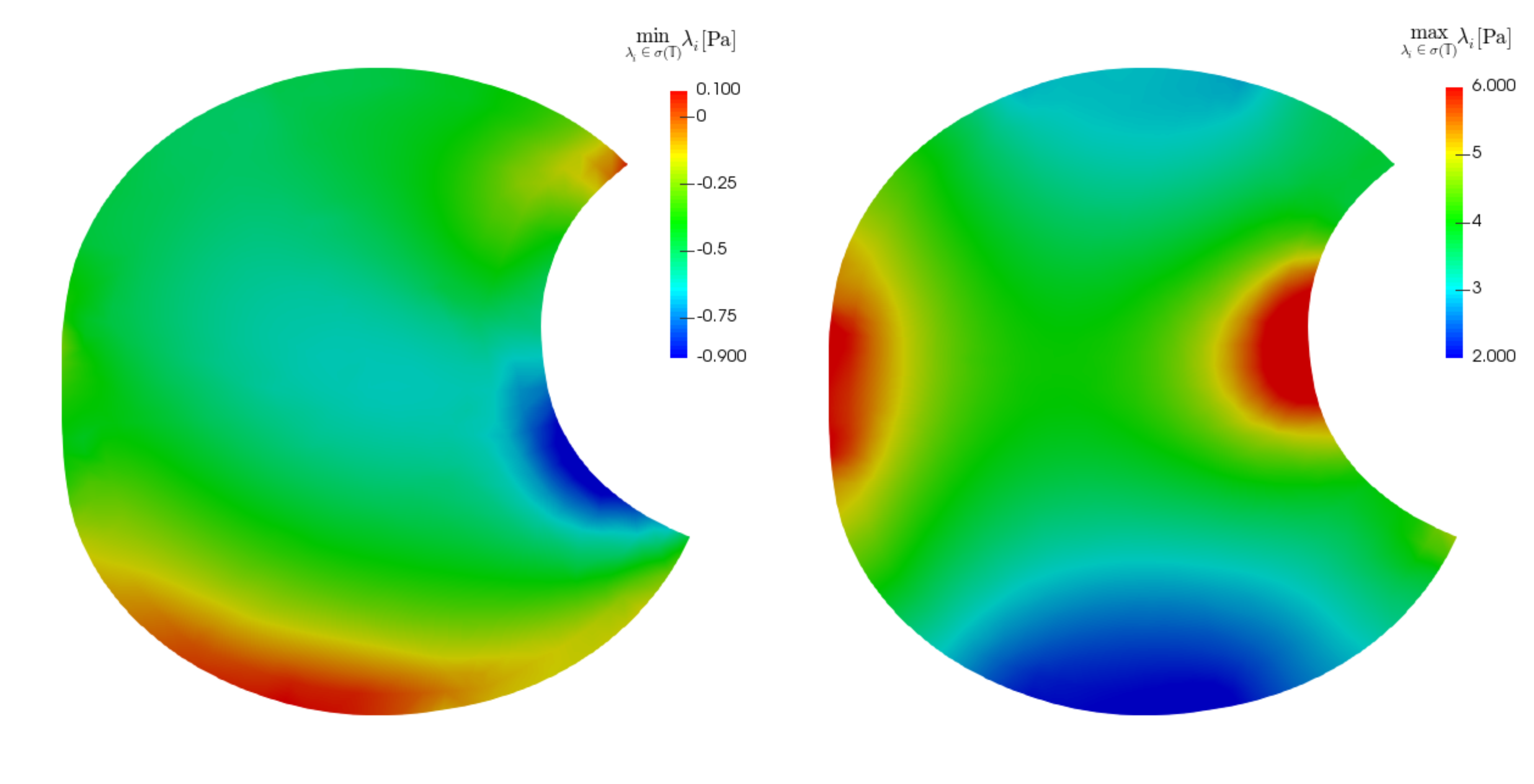}} 
\end{center}
\caption{
  Eigenvalues of the Cauchy stress tensor $\cstress$ at $t=\unit[120.5]{s}$ (half of the second prolongation), current configuration, cross-section with the plane $z=0$ and $z=\frac{h}{2}$. 
  Vitreous humour modelled using the Burgers-type viscoelastic model.}
\label{fig:stress-cross-section-eigenvalues-Burgers}
\end{figure}

\begin{figure}[h]
\begin{center}
\subfloat[Cross-section with the plane $z=0$.]{\includegraphics[width=0.8\textwidth]{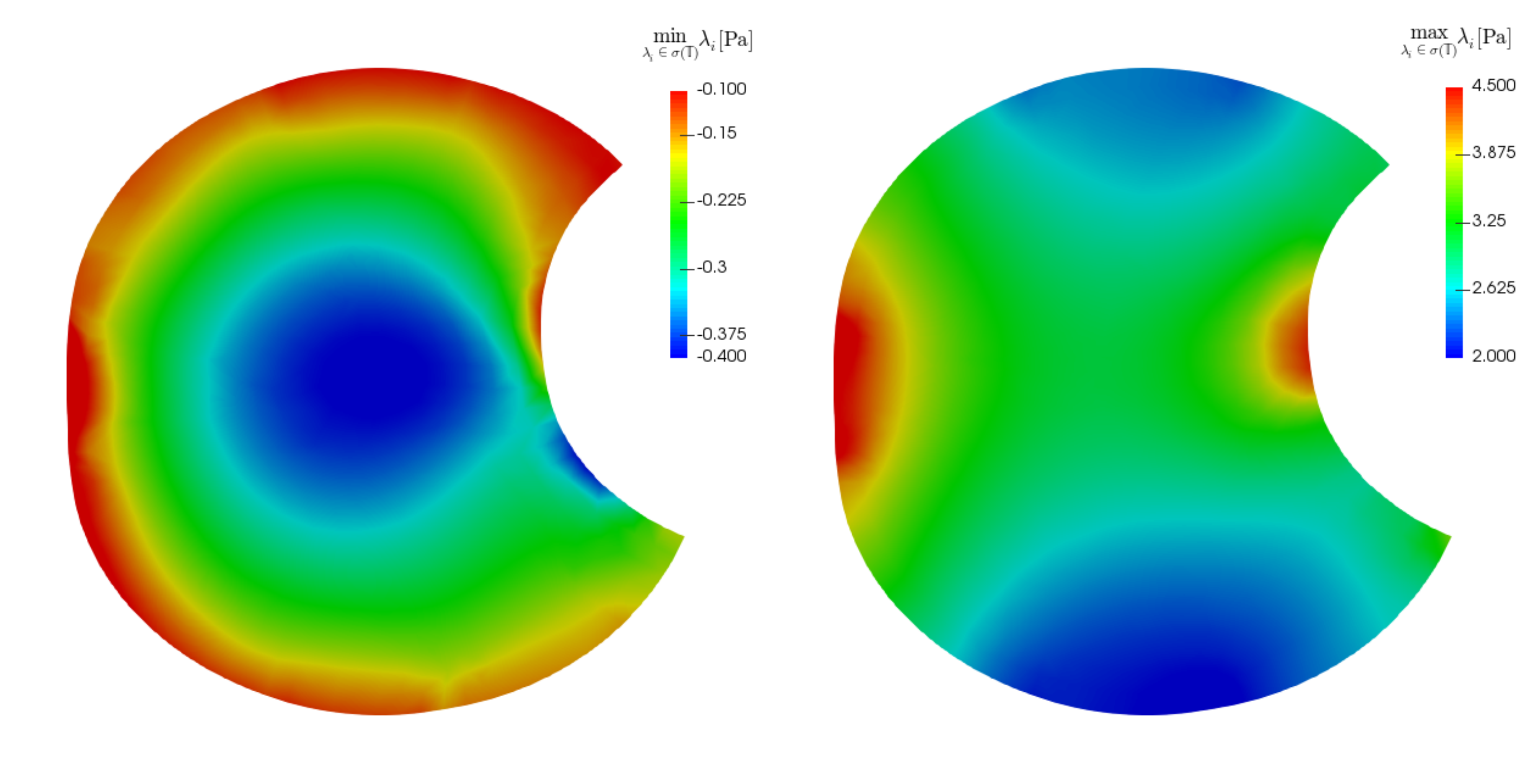}}\\
\subfloat[Cross-section with the plane $z=\frac{h}{2}$.]{\includegraphics[width=0.8\textwidth]{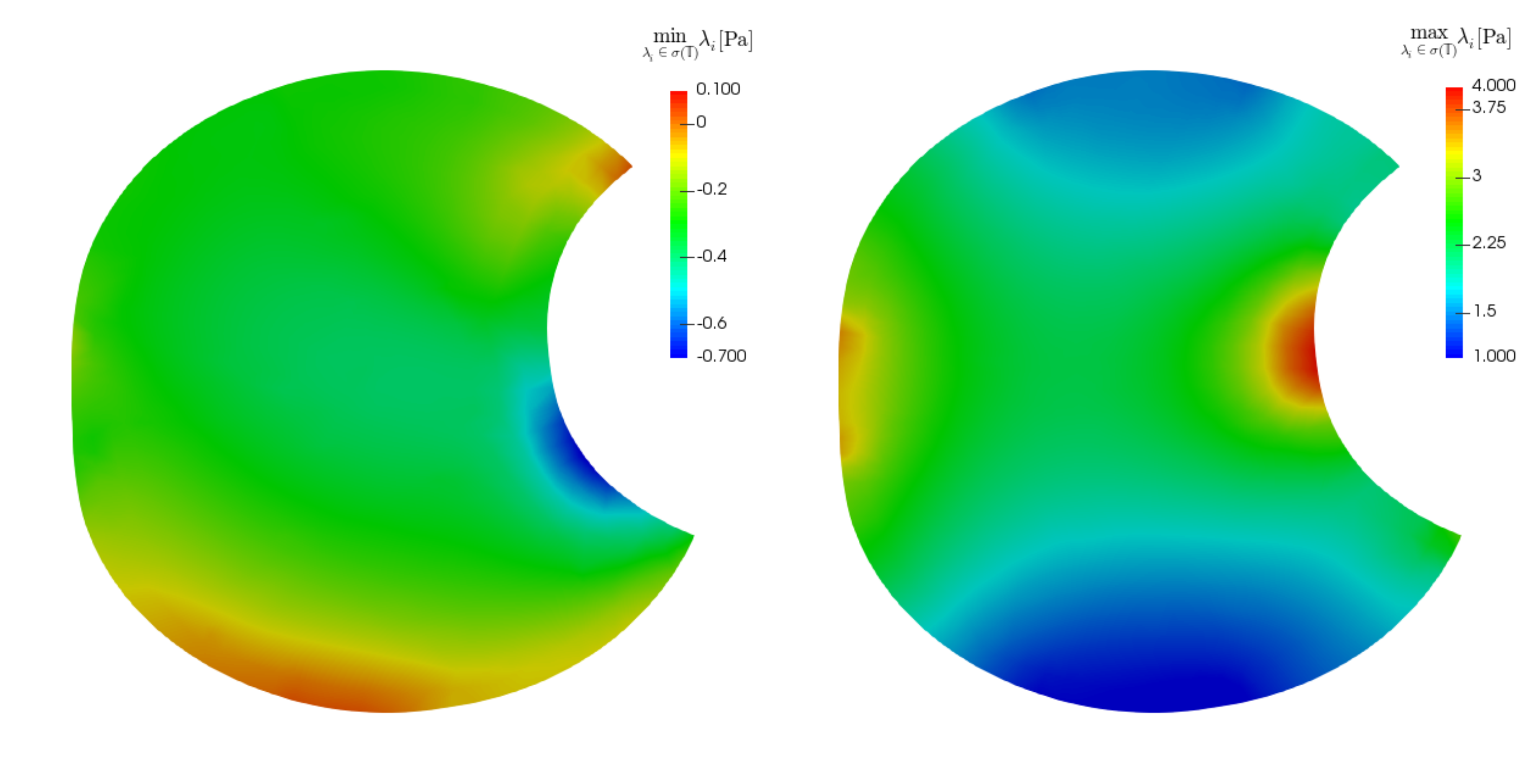}} 
\end{center}
\caption{
  Eigenvalues of the Cauchy stress tensor $\cstress$ at $t=\unit[120.5]{s}$ (half of the second prolongation), current configuration, cross-section with the plane $z=0$ and $z=\frac{h}{2}$. 
  Vitreous humour modelled using the Navier-Stokes model.}
\label{fig:stress-cross-section-eigenvalues-NS}
\end{figure}

Finally, in Figures~\ref{fig:stress-cross-section-eigenvalues-Burgers} and \ref{fig:stress-cross-section-eigenvalues-NS} we plot the spatial distribution of the minimal and maximal eigenvalues of the Cauchy stress tensor, which provides us a piece of information concerning the character of the stress at the given spatial point. (Positivity of the eigenvalue implies tension in the direction of the corresponding eigenvector, while the negativity of the eigenvalue implies compression in the direction of the corresponding eigenvector.)

%% file: conclusion.tex
A model for the deformation of an eyeball due to the application of a displacement on the surface of the eyeball has been presented. The model treats the eyeball as a cavity filled by a viscoelastic rate-type fluid (vitreous humour) that is enclosed by a hyperelastic solid (sclera, lens). The model allows one to study the motion of the vitreous humour and the displacement of the sclera and the lens, provided that the eyeball is subject to outer stimuli.

Moreover, besides the model we have also presented a numerical algorithm for solving the corresponding fluid-structure interaction problem. Since the numerical solution is based on the use of the finite element method, it in principle allows one to study the mechanical response of the eye in a patient specific geometry and in scenarios that are beyond the reach of analytical perturbation methods, see for example~\cite{ge.p.bottega.wj.ea:on}. The feasibility of the proposed numerical method has been demonstrated by the solution of an initial/boundary value problem for a set of realistic parameter values in a setting that resembles the recent experiment by~\cite{shah.ns.beebe.dc.ea:on}.

The main outcome of the trial computation is twofold. The overall flow pattern in the vitreous humour and the displacement field in the sclera and the lens seem to be, in the given scenario, not too much affected by the choice of the rheological model for the vitreous humour. On the other hand, the \emph{stress distribution in the vitreous humour and on the fluid/solid interface is very sensitive to the choice of the rheological properties of the vitreous humour}. Given the fact that some eye pathologies such as retinal detachment are thought to be closely linked to mechanical processes, see~\cite{kleinberg.tt.tzekov.rt.ea:vitreous}, the presented model can be possibly used to answer clinically relevant questions.






%% file: eye-computations-amsart.bbl
\begin{thebibliography}{43}
\providecommand{\natexlab}[1]{#1}
\providecommand{\url}[1]{\texttt{#1}}
\providecommand{\urlprefix}{URL }
\expandafter\ifx\csname urlstyle\endcsname\relax
  \providecommand{\doi}[1]{doi:\discretionary{}{}{}#1}\else
  \providecommand{\doi}[1]{doi:\discretionary{}{}{}\begingroup
  \urlstyle{rm}\url{#1}\endgroup}\fi
\providecommand{\bibinfo}[2]{#2}

\bibitem[{Kleinberg et~al.(2011)Kleinberg, Tzekov, Stein, Ravi, and
  Kaushal}]{kleinberg.tt.tzekov.rt.ea:vitreous}
\bibinfo{author}{T.~T. Kleinberg}, \bibinfo{author}{R.~T. Tzekov},
  \bibinfo{author}{L.~Stein}, \bibinfo{author}{N.~Ravi},
  \bibinfo{author}{S.~Kaushal}, \bibinfo{title}{Vitreous substitutes: {A}
  comprehensive review}, \bibinfo{journal}{Survey of Ophthalmology}
  \bibinfo{volume}{56}~(\bibinfo{number}{4}) (\bibinfo{year}{2011})
  \bibinfo{pages}{300--323},
  \doi{\bibinfo{doi}{10.1016/j.survophthal.2010.09.001}}.

\bibitem[{Shah et~al.(2016)Shah, Beebe, Lake, and
  Filas}]{shah.ns.beebe.dc.ea:on}
\bibinfo{author}{N.~S. Shah}, \bibinfo{author}{D.~C. Beebe},
  \bibinfo{author}{S.~P. Lake}, \bibinfo{author}{B.~A. Filas},
  \bibinfo{title}{On the spatiotemporal material anisotropy of the vitreous
  body in tension and compression}, \bibinfo{journal}{Ann. Biomed. Eng.}
  \bibinfo{volume}{44}~(\bibinfo{number}{10}) (\bibinfo{year}{2016})
  \bibinfo{pages}{3084--3095}, \doi{\bibinfo{doi}{10.1007/s10439-016-1589-3}}.

\bibitem[{David et~al.(1998)David, Smye, Dabbs, and
  James}]{david.t.smye.s.ea:model}
\bibinfo{author}{T.~David}, \bibinfo{author}{S.~Smye},
  \bibinfo{author}{T.~Dabbs}, \bibinfo{author}{T.~James}, \bibinfo{title}{A
  model for the fluid motion of vitreous humour of the human eye during
  saccadic movement}, \bibinfo{journal}{Phys. Med. Biol.}
  \bibinfo{volume}{43}~(\bibinfo{number}{6}) (\bibinfo{year}{1998})
  \bibinfo{pages}{1385--1399}, \doi{\bibinfo{doi}{10.1088/0031-9155/43/6/001}}.

\bibitem[{Repetto(2006)}]{repetto.r:analytical}
\bibinfo{author}{R.~Repetto}, \bibinfo{title}{An analytical model of the
  dynamics of the liquefied vitreous induced by saccadic eye movements},
  \bibinfo{journal}{Meccanica} \bibinfo{volume}{41}~(\bibinfo{number}{1})
  (\bibinfo{year}{2006}) \bibinfo{pages}{101--117},
  \doi{\bibinfo{doi}{10.1007/s11012-005-0782-5}}.

\bibitem[{Repetto et~al.(2010)Repetto, Siggers, and
  Stocchino}]{repetto.r.siggers.jh.ea:mathematical}
\bibinfo{author}{R.~Repetto}, \bibinfo{author}{J.~H. Siggers},
  \bibinfo{author}{A.~Stocchino}, \bibinfo{title}{Mathematical model of flow in
  the vitreous humor induced by saccadic eye rotations: effect of geometry},
  \bibinfo{journal}{Biomech. Model. Mechanobiol.}
  \bibinfo{volume}{9}~(\bibinfo{number}{1}) (\bibinfo{year}{2010})
  \bibinfo{pages}{65--76}, \doi{\bibinfo{doi}{10.1007/s10237-009-0159-0}}.

\bibitem[{Meskauskas et~al.(2011)Meskauskas, Repetto, and
  Siggers}]{meskauskas.j.repetto.r.ea:oscillatory}
\bibinfo{author}{J.~Meskauskas}, \bibinfo{author}{R.~Repetto},
  \bibinfo{author}{J.~H. Siggers}, \bibinfo{title}{Oscillatory motion of a
  viscoelastic fluid within a spherical cavity}, \bibinfo{journal}{J. Fluid
  Mech.} \bibinfo{volume}{685} (\bibinfo{year}{2011}) \bibinfo{pages}{1--22},
  \doi{\bibinfo{doi}{10.1017/jfm.2011.263}}.

\bibitem[{Meskauskas et~al.(2012)Meskauskas, Repetto, and
  Siggers}]{meskauskas.j.repetto.r.ea:shape}
\bibinfo{author}{J.~Meskauskas}, \bibinfo{author}{R.~Repetto},
  \bibinfo{author}{J.~H. Siggers}, \bibinfo{title}{Shape change of the vitreous
  chamber influences retinal detachment and reattachment processes: {I}s
  mechanical stress during eye rotations a factor?}, \bibinfo{journal}{Invest.
  Ophthalmol. Vis. Sci.} \bibinfo{volume}{53}~(\bibinfo{number}{10})
  (\bibinfo{year}{2012}) \bibinfo{pages}{6271--6281},
  \doi{\bibinfo{doi}{10.1167/iovs.11-9390}}.

\bibitem[{Bonfiglio et~al.(2013)Bonfiglio, Repetto, Siggers, and
  Stocchino}]{bonfiglio.a.repetto.r.ea:investigation}
\bibinfo{author}{A.~Bonfiglio}, \bibinfo{author}{R.~Repetto},
  \bibinfo{author}{J.~H. Siggers}, \bibinfo{author}{A.~Stocchino},
  \bibinfo{title}{Investigation of the motion of a viscous fluid in the
  vitreous cavity induced by eye rotations and implications for drug delivery},
  \bibinfo{journal}{Phys. Med. Biol.}
  \bibinfo{volume}{58}~(\bibinfo{number}{6}) (\bibinfo{year}{2013})
  \bibinfo{pages}{1969--1982},
  \doi{\bibinfo{doi}{10.1088/0031-9155/58/6/1969}}.

\bibitem[{Abouali et~al.(2012)Abouali, Modareszadeh, Ghaffariyeh, and
  Tu}]{abouali.o.modareszadeh.a.ea:numerical}
\bibinfo{author}{O.~Abouali}, \bibinfo{author}{A.~Modareszadeh},
  \bibinfo{author}{A.~Ghaffariyeh}, \bibinfo{author}{J.~Tu},
  \bibinfo{title}{Numerical simulation of the fluid dynamics in vitreous cavity
  due to saccadic eye movement}, \bibinfo{journal}{Medical Engineering \&
  Physics} \bibinfo{volume}{34}~(\bibinfo{number}{6}) (\bibinfo{year}{2012})
  \bibinfo{pages}{681--692},
  \doi{\bibinfo{doi}{10.1016/j.medengphy.2011.09.011}}.

\bibitem[{Modarreszadeh and
  Abouali(2014)}]{modarreszadeh.a.abouali.o:numerical}
\bibinfo{author}{A.~Modarreszadeh}, \bibinfo{author}{O.~Abouali},
  \bibinfo{title}{Numerical simulation for unsteady motions of the human
  vitreous humor as a viscoelastic substance in linear and non-linear regimes},
  \bibinfo{journal}{J. Non-Newtonian Fluid Mech.} \bibinfo{volume}{204}
  (\bibinfo{year}{2014}) \bibinfo{pages}{22--31},
  \doi{\bibinfo{doi}{10.1016/j.jnnfm.2013.12.001}}.

\bibitem[{Repetto et~al.(2014)Repetto, Siggers, and
  Meskauskas}]{repetto.r.siggers.jh.ea:steady}
\bibinfo{author}{R.~Repetto}, \bibinfo{author}{J.~H. Siggers},
  \bibinfo{author}{J.~Meskauskas}, \bibinfo{title}{Steady streaming of a
  viscoelastic fluid within a periodically rotating sphere},
  \bibinfo{journal}{J. Fluid Mech.} \bibinfo{volume}{761}
  (\bibinfo{year}{2014}) \bibinfo{pages}{329--347},
  \doi{\bibinfo{doi}{10.1017/jfm.2014.546}}.

\bibitem[{Lee et~al.(1992)Lee, Litt, and Buchsbaum}]{lee.b.litt.m.ea:rheology}
\bibinfo{author}{B.~Lee}, \bibinfo{author}{M.~Litt},
  \bibinfo{author}{G.~Buchsbaum}, \bibinfo{title}{Rheology of the vitreous
  body. {P}art {I}: Viscoelasticity of human vitreous},
  \bibinfo{journal}{Biorheology} \bibinfo{volume}{29}~(\bibinfo{number}{5--6})
  (\bibinfo{year}{1992}) \bibinfo{pages}{521--533}.

\bibitem[{Sharif-Kashani et~al.(2011)Sharif-Kashani, Hubschman, Sassoon, and
  Kavehpour}]{sharif-kashani.p.hubschman.j.ea:rheology}
\bibinfo{author}{P.~Sharif-Kashani}, \bibinfo{author}{J.-P. Hubschman},
  \bibinfo{author}{D.~Sassoon}, \bibinfo{author}{H.~P. Kavehpour},
  \bibinfo{title}{Rheology of the vitreous gel: {E}ffects of macromolecule
  organization on the viscoelastic properties}, \bibinfo{journal}{J. Biomech.}
  \bibinfo{volume}{44}~(\bibinfo{number}{3}) (\bibinfo{year}{2011})
  \bibinfo{pages}{419--423},
  \doi{\bibinfo{doi}{10.1016/j.jbiomech.2010.10.002}}.

\bibitem[{Giesekus(1982)}]{giesekus.h:simple}
\bibinfo{author}{H.~Giesekus}, \bibinfo{title}{A simple constitutive equation
  for polymer fluids based on the concept of deformation-dependent tensorial
  mobility}, \bibinfo{journal}{J. Non-Newton. Fluid Mech.}
  \bibinfo{volume}{11}~(\bibinfo{number}{1-2}) (\bibinfo{year}{1982})
  \bibinfo{pages}{69--109}, \doi{\bibinfo{doi}{10.1016/0377-0257(82)85016-7}}.

\bibitem[{Burgers(1939)}]{burgers.jm:mechanical}
\bibinfo{author}{J.~M. Burgers}, \bibinfo{title}{Mechanical considerations --
  Model systems -- Phenomenological theories of relaxation and viscosity}, in:
  \bibinfo{booktitle}{First report on viscosity and plasticity},
  chap.~\bibinfo{chapter}{1}, \bibinfo{publisher}{Nordemann Publishing},
  \bibinfo{address}{New York}, \bibinfo{pages}{5--67}, \bibinfo{year}{1939}.

\bibitem[{Rajagopal and
  Srinivasa(2000)}]{rajagopal.kr.srinivasa.ar:thermodynamic}
\bibinfo{author}{K.~R. Rajagopal}, \bibinfo{author}{A.~R. Srinivasa},
  \bibinfo{title}{A thermodynamic frame work for rate type fluid models},
  \bibinfo{journal}{J. Non-Newton. Fluid Mech.}
  \bibinfo{volume}{88}~(\bibinfo{number}{3}) (\bibinfo{year}{2000})
  \bibinfo{pages}{207--227},
  \doi{\bibinfo{doi}{10.1016/S0377-0257(99)00023-3}}.

\bibitem[{Donea et~al.(2004)Donea, Huerta, Ponthot, and
  Rodr\'{\i}guez-Ferrari}]{donea.j.huerta.a.ea:arbitrary}
\bibinfo{author}{J.~Donea}, \bibinfo{author}{A.~Huerta}, \bibinfo{author}{J.-P.
  Ponthot}, \bibinfo{author}{Rodr\'{\i}guez-Ferrari}, \bibinfo{title}{Arbitrary
  {L}agrangian--{E}ulerian methods}, in: \bibinfo{booktitle}{Encyclopedia of
  Computational Mechanics}, \bibinfo{publisher}{John Wiley \& Sons}, ISBN
  \bibinfo{isbn}{9780470091357},
  \doi{\bibinfo{doi}{10.1002/0470091355.ecm009}}, \bibinfo{year}{2004}.

\bibitem[{Geuzaine and Remacle(2009)}]{geuzaine.c.remacle.j:gmsh}
\bibinfo{author}{C.~Geuzaine}, \bibinfo{author}{J.-F. Remacle},
  \bibinfo{title}{Gmsh: A 3-{D} finite element mesh generator with built-in
  pre- and post-processing facilities}, \bibinfo{journal}{Int. J. Numer.
  Methods Eng.} \bibinfo{volume}{79}~(\bibinfo{number}{11})
  (\bibinfo{year}{2009}) \bibinfo{pages}{1309--1331},
  \doi{\bibinfo{doi}{10.1002/nme.2579}}.

\bibitem[{Wineman and Rajagopal(2000)}]{wineman.as.rajagopal.kr:mechanical}
\bibinfo{author}{A.~S. Wineman}, \bibinfo{author}{K.~R. Rajagopal},
  \bibinfo{title}{Mechanical response of polymers---an introduction},
  \bibinfo{publisher}{Cambridge University Press},
  \bibinfo{address}{Cambridge}, \bibinfo{year}{2000}.

\bibitem[{Karra and Rajagopal(2009)}]{karra.s.rajagopal.kr:development}
\bibinfo{author}{S.~Karra}, \bibinfo{author}{K.~R. Rajagopal},
  \bibinfo{title}{Development of three dimensional constitutive theories based
  on lower dimensional experimental data}, \bibinfo{journal}{Appl. Mat.}
  \bibinfo{volume}{54}~(\bibinfo{number}{2}) (\bibinfo{year}{2009})
  \bibinfo{pages}{147--176}, \doi{\bibinfo{doi}{10.1007/s10492-009-0010-z}}.

\bibitem[{Rao and Rajagopal(2002)}]{rao.ij.rajagopal.kr:thermodynamic}
\bibinfo{author}{I.~J. Rao}, \bibinfo{author}{K.~R. Rajagopal},
  \bibinfo{title}{A thermodynamic framework for the study of crystallization in
  polymers}, \bibinfo{journal}{Z. Angew. Math. Phys.}
  \bibinfo{volume}{53}~(\bibinfo{number}{3}) (\bibinfo{year}{2002})
  \bibinfo{pages}{365--406}, \doi{\bibinfo{doi}{10.1007/s00033-002-8161-8}}.

\bibitem[{Sodhi and Rao(2010)}]{sodhi.js.rao.ij:modeling}
\bibinfo{author}{J.~S. Sodhi}, \bibinfo{author}{I.~J. Rao},
  \bibinfo{title}{Modeling the mechanics of light activated shape memory
  polymers}, \bibinfo{journal}{Int. J. Eng. Sci.}
  \bibinfo{volume}{48}~(\bibinfo{number}{11}) (\bibinfo{year}{2010})
  \bibinfo{pages}{1576--1589},
  \doi{\bibinfo{doi}{10.1016/j.ijengsci.2010.05.003}}.

\bibitem[{M\'alek et~al.(2015{\natexlab{a}})M\'alek, Rajagopal, and
  T\r{u}ma}]{malek.j.rajagopal.kr.ea:on}
\bibinfo{author}{J.~M\'alek}, \bibinfo{author}{K.~R. Rajagopal},
  \bibinfo{author}{K.~T\r{u}ma}, \bibinfo{title}{On a variant of the {M}axwell
  and {O}ldroyd-{B} models within the context of a thermodynamic basis},
  \bibinfo{journal}{Int. J. Non-Linear Mech.} \bibinfo{volume}{76}
  (\bibinfo{year}{2015}{\natexlab{a}}) \bibinfo{pages}{42--47},
  \doi{\bibinfo{doi}{10.1016/j.ijnonlinmec.2015.03.009}}.

\bibitem[{\v{R}eho\v{r} et~al.(2016)\v{R}eho\v{r}, Pr\r{u}\v{s}a, and
  T\r{u}ma}]{rehor.m.pusa.v.ea:on}
\bibinfo{author}{M.~\v{R}eho\v{r}}, \bibinfo{author}{V.~Pr\r{u}\v{s}a},
  \bibinfo{author}{K.~T\r{u}ma}, \bibinfo{title}{On the response of nonlinear
  viscoelastic materials in creep and stress relaxation experiments in the
  lubricated squeeze flow setting}, \bibinfo{journal}{Phys. Fluids}
  \bibinfo{volume}{28}~(\bibinfo{number}{10}) \bibinfo{eid}{103102},
  \doi{\bibinfo{doi}{10.1063/1.4964662}}.

\bibitem[{Kwack et~al.(2016)Kwack, Masud, and
  Rajagopal}]{kwack.j.masud.a.ea:stabilized}
\bibinfo{author}{J.~Kwack}, \bibinfo{author}{A.~Masud}, \bibinfo{author}{K.~R.
  Rajagopal}, \bibinfo{title}{Stabilized mixed three-field formulation for a
  generalized incompressible {O}ldroyd-{B} model}, \bibinfo{journal}{Int. J.
  Numer. Meth. Fluids} \bibinfo{volume}{83} (\bibinfo{year}{2016})
  \bibinfo{pages}{704--734}, \doi{\bibinfo{doi}{10.1002/fld.4287}}.

\bibitem[{Hron et~al.(2017)Hron, Milo\v{s}, Pr\r{u}\v{s}a, Sou\v{c}ek, and
  T\r{u}ma}]{hron.j.milos.v.ea:on}
\bibinfo{author}{J.~Hron}, \bibinfo{author}{V.~Milo\v{s}},
  \bibinfo{author}{V.~Pr\r{u}\v{s}a}, \bibinfo{author}{O.~Sou\v{c}ek},
  \bibinfo{author}{K.~T\r{u}ma}, \bibinfo{title}{On thermodynamics of
  viscoelastic rate type fluids with temperature dependent material
  coefficients}, \bibinfo{journal}{Int. J. Non-Linear Mech.}
  \bibinfo{volume}{95} (\bibinfo{year}{2017}) \bibinfo{pages}{193--208},
  \doi{\bibinfo{doi}{10.1016/j.ijnonlinmec.2017.06.011}}.

\bibitem[{T\r{u}ma(2013)}]{tuma.k:identification}
\bibinfo{author}{K.~T\r{u}ma}, \bibinfo{title}{Identification of rate type
  fluids suitable for modeling geomaterials}, Ph.D. thesis,
  \bibinfo{school}{Charles University}, \bibinfo{year}{2013}.

\bibitem[{Hron et~al.(2014)Hron, Rajagopal, and
  T\r{u}ma}]{hron.j.rajagopal.kr.ea:flow}
\bibinfo{author}{J.~Hron}, \bibinfo{author}{K.~R. Rajagopal},
  \bibinfo{author}{K.~T\r{u}ma}, \bibinfo{title}{Flow of a {B}urgers fluid due
  to time varying loads on deforming boundaries}, \bibinfo{journal}{J.
  Non-Newton. Fluid Mech.} \bibinfo{volume}{210} (\bibinfo{year}{2014})
  \bibinfo{pages}{66--77}, \doi{\bibinfo{doi}{10.1016/j.jnnfm.2014.05.005}}.

\bibitem[{M\'alek et~al.(2015{\natexlab{b}})M\'alek, Rajagopal, and
  T\r{u}ma}]{malek.j.rajagopal.kr.ea:thermodynamically}
\bibinfo{author}{J.~M\'alek}, \bibinfo{author}{K.~R. Rajagopal},
  \bibinfo{author}{K.~T\r{u}ma}, \bibinfo{title}{A thermodynamically compatible
  model for describing the response of asphalt binders}, \bibinfo{journal}{Int.
  J. Pavement Eng.} \bibinfo{volume}{16}~(\bibinfo{number}{4})
  (\bibinfo{year}{2015}{\natexlab{b}}) \bibinfo{pages}{297--314},
  \doi{\bibinfo{doi}{10.1080/10298436.2014.942860}}.

\bibitem[{M\'alek et~al.(2016)M\'alek, Rajagopal, and
  T\r{u}ma}]{malek.j.rajagopal.kr.ea:thermodynamically*1}
\bibinfo{author}{J.~M\'alek}, \bibinfo{author}{K.~R. Rajagopal},
  \bibinfo{author}{K.~T\r{u}ma}, \bibinfo{title}{A thermodynamically compatible
  model for describing asphalt binders: solutions of problems},
  \bibinfo{journal}{Int. J. Pavement Eng.}
  \bibinfo{volume}{17}~(\bibinfo{number}{6}) (\bibinfo{year}{2016})
  \bibinfo{pages}{550--564},
  \doi{\bibinfo{doi}{10.1080/10298436.2015.1007575}}.

\bibitem[{Narayan et~al.(2016)Narayan, Little, and
  Rajagopal}]{narayan.spa.little.dn.ea:modelling}
\bibinfo{author}{S.~P.~A. Narayan}, \bibinfo{author}{D.~N. Little},
  \bibinfo{author}{K.~R. Rajagopal}, \bibinfo{title}{Modelling the nonlinear
  viscoelastic response of asphalt binders}, \bibinfo{journal}{Int. J. Pavement
  Eng.} \bibinfo{volume}{17}~(\bibinfo{number}{2}) (\bibinfo{year}{2016})
  \bibinfo{pages}{123--132}, \doi{\bibinfo{doi}{10.1080/10298436.2014.925621}}.

\bibitem[{Murphy et~al.(2016)Murphy, Black, and Hastings}]{black2013handbook}
\bibinfo{editor}{W.~Murphy}, \bibinfo{editor}{J.~Black},
  \bibinfo{editor}{G.~Hastings} (Eds.), \bibinfo{title}{Handbook of Biomaterial
  Properties}, \bibinfo{publisher}{Springer},
  \doi{\bibinfo{doi}{10.1007/978-1-4939-3305-1}}, \bibinfo{year}{2016}.

\bibitem[{Grytz et~al.(2014)Grytz, Fazio, Girard, Libertiaux, Bruno, Gardiner,
  Girkin, and Downs}]{grytz.r.fazio.ma.ea:material}
\bibinfo{author}{R.~Grytz}, \bibinfo{author}{M.~A. Fazio},
  \bibinfo{author}{M.~J. Girard}, \bibinfo{author}{V.~Libertiaux},
  \bibinfo{author}{L.~Bruno}, \bibinfo{author}{S.~Gardiner},
  \bibinfo{author}{C.~A. Girkin}, \bibinfo{author}{J.~C. Downs},
  \bibinfo{title}{Material properties of the posterior human sclera},
  \bibinfo{journal}{J. Mech. Behav. Biomed. Mater.} \bibinfo{volume}{29}
  (\bibinfo{year}{2014}) \bibinfo{pages}{602--617},
  \doi{\bibinfo{doi}{10.1016/j.jmbbm.2013.03.027}}.

\bibitem[{Wilde et~al.(2012)Wilde, Burd, and Judge}]{wilde.gs.burd.hj.ea:shear}
\bibinfo{author}{G.~S. Wilde}, \bibinfo{author}{H.~J. Burd},
  \bibinfo{author}{S.~J. Judge}, \bibinfo{title}{Shear modulus data for the
  human lens determined from a spinning lens test}, \bibinfo{journal}{Exp. Eye
  Res.} \bibinfo{volume}{97}~(\bibinfo{number}{1}) (\bibinfo{year}{2012})
  \bibinfo{pages}{36--48}, \doi{\bibinfo{doi}{10.1016/j.exer.2012.01.011}}.

\bibitem[{Su et~al.(2009)Su, Vesco, Fleming, and Choh}]{dens}
\bibinfo{author}{X.~Su}, \bibinfo{author}{C.~Vesco},
  \bibinfo{author}{J.~Fleming}, \bibinfo{author}{V.~Choh},
  \bibinfo{title}{Density of ocular components of the bovine eye},
  \bibinfo{journal}{Optom. Vis. Sci.}
  \bibinfo{volume}{86}~(\bibinfo{number}{10}) (\bibinfo{year}{2009})
  \bibinfo{pages}{1187--1195},
  \doi{\bibinfo{doi}{10.1097/OPX.0b013e3181baaf4e}}.

\bibitem[{Truesdell and Noll(2004)}]{truesdell.c.noll.w:non-linear*1}
\bibinfo{author}{C.~Truesdell}, \bibinfo{author}{W.~Noll}, \bibinfo{title}{The
  non-linear field theories of mechanics}, \bibinfo{publisher}{Springer},
  \bibinfo{address}{Berlin}, \bibinfo{edition}{3rd} edn., \bibinfo{year}{2004}.

\bibitem[{Scovazzi and Hughes(2007)}]{scovazzi.g.hughes.t:lecture}
\bibinfo{author}{G.~Scovazzi}, \bibinfo{author}{T.~Hughes},
  \bibinfo{title}{Lecture notes on continuum mechanics on arbitrary moving
  domains}, \bibinfo{type}{Tech. Rep.}, \bibinfo{institution}{Technical Report
  SAND-2007-6312P, Sandia National Laboratories}, \bibinfo{year}{2007}.

\bibitem[{Razzaq et~al.(2010)Razzaq, Hron, and
  Turek}]{razzaq.m.hron.j.ea:numerical}
\bibinfo{author}{M.~Razzaq}, \bibinfo{author}{J.~Hron},
  \bibinfo{author}{S.~Turek}, \bibinfo{title}{Numerical simulation of laminar
  incompressible fluid-structure interaction for elastic material with point
  constraints}, in: \bibinfo{booktitle}{Advances in mathematical fluid
  mechanics}, \bibinfo{publisher}{Springer}, \bibinfo{address}{Berlin},
  \bibinfo{pages}{451--472}, \doi{\bibinfo{doi}{10.1007/978-3-642-04068-9_27}},
  \bibinfo{year}{2010}.

\bibitem[{Hron and M\'adl\'{\i}k(2007)}]{hron.j.madlik.m:fluid-structure}
\bibinfo{author}{J.~Hron}, \bibinfo{author}{M.~M\'adl\'{\i}k},
  \bibinfo{title}{Fluid-structure interaction with applications in
  biomechanics}, \bibinfo{journal}{Nonlinear Anal.-Real World Appl.}
  \bibinfo{volume}{8}~(\bibinfo{number}{5}) (\bibinfo{year}{2007})
  \bibinfo{pages}{1431--1458},
  \doi{\bibinfo{doi}{10.1016/j.nonrwa.2006.05.007}}.

\bibitem[{Korelc(2002)}]{korelc.j:Multi}
\bibinfo{author}{J.~Korelc}, \bibinfo{title}{Multi-language and
  multi-environment generation of nonlinear finite element codes},
  \bibinfo{journal}{Eng. Comput.} \bibinfo{volume}{18}~(\bibinfo{number}{4})
  (\bibinfo{year}{2002}) \bibinfo{pages}{312--327},
  \doi{\bibinfo{doi}{10.1007/s003660200028}}.

\bibitem[{Korelc(2009)}]{korelc.j:Automation}
\bibinfo{author}{J.~Korelc}, \bibinfo{title}{Automation of primal and
  sensitivity analysis of transient coupled problems}, \bibinfo{journal}{Comp.
  Mech.} \bibinfo{volume}{44}~(\bibinfo{number}{5}) (\bibinfo{year}{2009})
  \bibinfo{pages}{631--649}, \doi{\bibinfo{doi}{10.1007/s00466-009-0395-2}}.

\bibitem[{Sonneveld(1989)}]{sonneveld.p:CGS}
\bibinfo{author}{P.~Sonneveld}, \bibinfo{title}{{CGS}, {A} fast {L}anczos-type
  solver for nonsymmetric linear systems}, \bibinfo{journal}{SIAM J. Sci. Stat.
  Comput.} \bibinfo{volume}{10}~(\bibinfo{number}{1}) (\bibinfo{year}{1989})
  \bibinfo{pages}{36--52}, \doi{\bibinfo{doi}{10.1137/0910004}}.

\bibitem[{Ge et~al.(2017)Ge, Bottega, Prenner, and
  Fine}]{ge.p.bottega.wj.ea:on}
\bibinfo{author}{P.~Ge}, \bibinfo{author}{W.~J. Bottega},
  \bibinfo{author}{J.~L. Prenner}, \bibinfo{author}{H.~F. Fine},
  \bibinfo{title}{On the behavior of an eye encircled by a scleral buckle},
  \bibinfo{journal}{J. Math. Biol.} \bibinfo{volume}{74}~(\bibinfo{number}{1})
  (\bibinfo{year}{2017}) \bibinfo{pages}{313--332},
  \doi{\bibinfo{doi}{10.1007/s00285-016-1015-3}}.

\end{thebibliography}
